\documentclass[letterpaper, 10 pt, conference]{ieeeconf}  % Comment this line out
\IEEEoverridecommandlockouts                              % This command is only
\title{\LARGE \bf
	State estimation of an electrochemical lithium-ion battery model: improved observer performance by hybrid redesign*
}

\author{E. Petri$^{1}$, T. Reynaudo$^{1}$,	R. Postoyan$^{1}$, D. Astolfi$^{2}$, D. Ne\v{s}i\'c$^{3}$ and S. Ra\"{e}l$^{4}$ 
	\thanks{*This work was funded by Lorraine Universit\'e d'Excellence LUE, HANDY project ANR-18-CE40-0010-02, the France Australia collaboration project IRP-ARS CNRS and the Australian Research Council under the Discovery Project DP200101303.  }
	\thanks{$^{1}$
		E. Petri, T. Reynaudo and R. Postoyan are with the Universit\'e de Lorraine, CNRS, CRAN, F-54000 Nancy, France.   
		(\tt\footnotesize elena.petri@univ-lorraine.fr).} %, thomas.reynaudo8@etu.univ-lorraine.fr, romain.postoyan@univ-lorraine.fr).}
	\thanks{$^{2}$D. Astolfi is with  
		Universit\'e Claude Bernard Lyon 1, CNRS, LAGEPP UMR 5007, F-69100,
		Villeurbanne, France.} % (\tt\footnotesize daniele.astolfi@univ-lyon1.fr).}
	\thanks{$^{3}$ D. Ne\v{s}i\'c is with the Department of Electrical and Electronic Engineering, The University of Melbourne, Parkville, 3010 Victoria Australia.}
\thanks{$^{4}$S. Ra\"{e}l is with	Université de Lorraine, GREEN, F-54000 Nancy, France.} % (\tt\footnotesize stephane.rael@univ-lorraine.fr).}
}

\usepackage{graphicx,fleqn,amstext,amsmath,amssymb,color,psfrag,mathabx,booktabs}
\usepackage[latin1]{inputenc}
\usepackage[colorlinks=true, letterpaper,    %no frame around URL
urlcolor=black,     %no colors
citecolor=blue,
menucolor=black,    %no colors
linkcolor=black,    %no colors
pagecolor=black,    %no colors
breaklinks=true,
]{hyperref}

\usepackage{amsthm}

\usepackage[noadjust]{cite}

\usepackage{comment}

\newcommand{\R}{\ensuremath{\mathbb{R}}}

\newcommand{\Rlo}{\ensuremath{\mathbb{R}_{\geq 0}}}

\newcommand{\Zo}{\ensuremath{\mathbb{Z}_{\geq 0}}}
\newcommand{\Zp}{\ensuremath{\mathbb{Z}_{> 0}}}
\newcommand{\Z}{\ensuremath{\mathbb{Z}}}

\definecolor{bleucit}{rgb}{0.2,0.4,0.6} % nouvelle couleur

\definecolor{blue_cv}{rgb}{0.09,0.35,0.78}

\newcommand{\KL}{\ensuremath{\mathcal{KL}}}
\newcommand{\K}{\ensuremath{\mathcal{K}}}
\newcommand{\Kinf}{\ensuremath{\mathcal{K}_{\infty}}}

\newcommand{\norm}[1]{\ensuremath{\left\|{#1}\right\|}}

\theoremstyle{theorem}

\newtheorem{ass}{\textnormal{\textbf{Assumption}}}

\newtheorem{thm}{\textnormal{\textbf{Theorem}}}

\usepackage{array}

\usepackage{mathtools}

\usepackage[noadjust]{cite}

\let\labelindent\relax
\usepackage{enumitem}
\usepackage{tikz}
\usetikzlibrary{shapes,arrows,calc, decorations.pathreplacing,calligraphy}
\definecolor{MyGreen}{RGB}{50,140,80}

\usepackage[export]{adjustbox}
\usepackage{ circuitikz }
\usetikzlibrary{arrows}

\usepackage{algorithm}
\usepackage{algpseudocode}

\usepackage{multirow}

\usetikzlibrary{backgrounds}

\begin{document}

	\maketitle
	\thispagestyle{empty}
	\pagestyle{empty}

	%%%%%%%%%%%%%%%%%%%%%%%%%%%%%%%%%%%%%%%%%%%%%%%%%%%%%%%%%%%%%%%%%%%%%%%%%%%%%%%%
	\begin{abstract}
	% Lithium-ion batteries are one of the most used energy storage technologies. %thanks to their good properties.
	 Effective management and just-in-time maintenance of lithium-ion batteries require the knowledge of unmeasured (internal) variables that need to be estimated.
	 %However, some reliable information of the internal state of the battery, that is not available throght measurement, is required for their usage. 
	 Observers are thus designed for this purpose using a mathematical model of the battery internal dynamics. It appears that it is often difficult to tune the observers to obtain good estimation performances both  in terms of convergence speed and accuracy, while these are essential in practice.
	 In this context, we demonstrate how a recently developed hybrid multi-observer can be used to improve the performance of a given observer designed for an electrochemical model of a lihium-ion battery.
	 %
	  %in the literature based on an electrochemical battery model. 
	% However, it is very hard to tune their gains to obtain good estimation performance.
%	  For this reason we apply, and extend, the hybrid multi-observer presented in \cite{petri2022towards} to improve the estimation performance of a given nominal observer. In this paper, the nominal observer that estimates the lithium concentrations of the elctrochemical battery, is the one proposed in \cite{blondel2017observer}, whose design is based on a polytopic approach. % and it is used to %, designed in the literature to
	   % For this reason we apply, and extend, the hybrid multi-observer presented in \cite{petri2022towards} to improve the estimation performance of the observer proposed in \cite{blondel2017observer}, whose design is based on a polytopic approach. %, and that estimates the lithium concentrations of the elctrochemical battery.   % and it is used to %, designed in the literature to
	 Simulation results, obtained with standard parameters values, show the estimation performance improvement using the proposed method. % in the context of battery estimation. 		
	\end{abstract}

%%%%%%%%%%%%%%%%%%%%%%%%%%%%%%%%%%%%%%%%%%%%%%%%%%%%%%%%%%%%%%%%%%%%%%%%%%%%%%%%
\section{INTRODUCTION}
%\textcolor{red}{To be written}
%\color{blue}
Lithium-ion batteries are widely used for the many advantages they exhibit in terms of volume capacity, weight, power density and the absence of %\textcolor{blue}{are less affected by} 
memory effect, compared to other energy storage technologies. On the other hand, the so-called battery management system (BMS) is required for a safe and efficient usage of the battery. The BMS impacts the battery performance and lifespan and it depends on the actual state of charge (SOC) of the battery, which is directly related to the lithium concentrations in the battery electrodes. An accurate knowledge of the SOC is therefore essential for proper battery management. Unfortunately, the SOC cannot be measured directly %as well as the lithium concentration in the electrodes, 
and thus needs to be estimated from the measured variables, typically the current and the voltage.  To address this challenge, a common approach is to design observers, based on a mathematical model of the internal dynamics, to estimate the unmeasured internal states, see e.g., \cite{he2012comparison, meng2018overview}. This task is non-trivial because of the nonlinear relationships between the internal variables and the measured ones.
%\color{blue}
Several approaches are available in the literature depending on the type of battery model (equivalent circuit model, infinite/finite-dimensional electrochemical models) and the type of observers, see, e.g.,~\cite{lee2008state,barillas2015comparative,xia2014novel, doyle1993modeling, fuller1994simulation,di2010lithium, blondel2017observer, blondel2018nonlinear,plante2022multiple}.

In this work, we focus on the finite-dimensional electrochemical model considered in \cite{blondel2017observer, blondel2018nonlinear, rael2013using}, which is derived from the infinite-dimensional models in  \cite{doyle1993modeling, fuller1994simulation}, as it offers a good compromise between accuracy and computational complexity.
%\color{black}
%\begin{comment}
% A common approach consists in using electrochemical models, which are based on partial differential equations used to describe the lithium-ion battery dynamics, see e.g., \cite{doyle1993modeling, fuller1994simulation} for the first works in this field. The electrochemical battery model in these works is very complex and it was simplified, considering each electrode as a spherical particle, in \cite{di2010lithium} where a single particle model (SPM) was presented. Moreover, to further simplify the electrochemical model, the partial differential equations have been discretized in e.g., \cite{blondel2017observer, blondel2018nonlinear,plante2022multiple}, where the lithium diffusion in each electrode is described by a set of ordinary differential equations. In this paper we consider the eletrochemical lithium-ion battery model proposed in \cite{blondel2018nonlinear} because it is simpler than ones described by partial differential equations and it has been shown that it is an accurate model for the battery. In particular, 
% \end{comment}
 The model takes the form of an affine system with a nonlinear output map, where the system states are the lithium concentrations in the electrodes, the input is the current and the measured output is the voltage. 
% \color{blue}
 A globally convergent observer was designed for this model in \cite{blondel2017observer} based on a polytopic approach. The issue is that to tune this observer to obtain both fast convergence and good robustness properties with respect to measurement noise and model uncertainties is highly non-trivial. %\textcolor{blue}{
 	The objective of this work is to address this challenge by systematically improving the estimation performance of an observer designed as in \cite{blondel2017observer} using a multi-observer approach (see, e.g., \cite[Section 8.3]{bernard2022observer}). In particular, we follow the hybrid methodology we recently developed in \cite{petri2022towards}, which 
% \color{black}
%Considering this model, in \cite{blondel2017observer} an observer based on a polytopic approach was designed to estimate the lithium concentration, and consequently, have an estimate of the SOC. Similarly, in \cite{blondel2018nonlinear} a nonlinear circle criterion observer was proposed. 
%The observer designs proposed in e.g., \cite{blondel2017observer, blondel2018nonlinear} guarantee a convergence property of the estimation error, however, it is really difficult to tune their gains to obtain good performance in terms of convergence speed and noise and disturbance rejection. Moreover, the conditions used to obtain the gains are typically very conservative. 
%
%The objective of this paper is to improve the estimation performance of the observer presented in \cite{blondel2017observer} applying the hybrid multi-observer proposed in \cite{petri2022towards} to the context of electrochemical lithium-ion battery estimation. 
%In this paper we apply the hybrid multi-observer proposed in \cite{petri2022towards} to the contest of electrochemical lithium-ion battery estimation with the aim of improving the estimation performance of the observer presented in \cite{blondel2017observer}. 
%The methodology in \cite{petri2022towards} 
consists in first designing a nominal observer using \cite{blondel2017observer} that satisfies an input-to-state stability property. %} %In this paper, we use the observer designed in \cite{blondel2017observer} as a starting point (nominal observer) in applying results in \cite{petri2022towards}.
Then, a bank of additional observer-like systems, that differ from the nominal one only on their gains, are added in parallel to the nominal observer. The gains of these additional dynamical systems can be arbitrary selected and do not need to be tuned to guarantee a convergence property of their estimation errors. These gains can thus be selected using any analytical or heuristic method to improve the convergence speed or the robustness of the nominal observer. Each of these systems, as well as the nominal observer, is called mode for the sake of convenience. To evaluate the performance of each mode, monitoring variables are introduced.
%
%Some monitoring variables are then used to evaluate the performance of each mode of the multi-observer. 
Based on these monitoring variables, the ``best mode'' is then selected at any time instant and its state estimate is considered for the battery internal state estimation. %is selected at any time instant.
 Therefore, the state estimate of the hybrid multi-observer switches between the states estimates of the modes and thus it is called hybrid.  The observer is modeled as an hybrid system using the formalism of \cite{goebel2012hybrid}. Note that, due to these switching, the state estimate exhibits discontinuities, which can be a problem for batteries, as this means the SOC estimate would experience jumps. For this reason, in this work, we add a filtered version of the hybrid multi-observer state estimate to the observer presented in \cite{petri2022towards}. 
%\color{blue}
 We provide an input-to-state stability property with respect to measurement noise, perturbation and disturbance for the new hybrid system and, as in \cite{petri2022towards}, we show that the performance of the hybrid multi-observer is, at least, as good as the performance of the nominal observer \cite{blondel2017observer}. 
 To illustrate the efficiency of the hybrid scheme, we present simulation results where a higher fidelity model of the battery is used to generate the output voltage compared to the one used to design the observers. Using the technique in \cite{blondel2017observer}, we first design the nominal observer, which shows good transient performance in terms of speed and small overshoot, but whose accuracy in steady-state may not be satisfactory. To address this issue, we select the gains of the additional modes of the hybrid multi-observer \cite{petri2022towards} smaller than the nominal one, with the aim of improving the robustness to noises and perturbations.   Simulation results show that the estimation performance are significantly improved with the hybrid multi-observer \cite{petri2022towards}, thereby illustrating the potential of this approach.
 % \color{black}
% 
%  Simulation results on a higher dimensional, and thus higher fidelity, electrochemical model, for standard parameters values are finally presented to illustrate that the hybrid multi-observer \cite{petri2022towards} improves the estimation performance of the nominal observer in \cite{blondel2017observer}. 

% \cite{goebel2012hybrid} and a multi-observer

%\color{black}
\begin{comment}
The reminder of the paper is organized as follows. In Section~\ref{BatteryModel} we recall the electrochemical battery model proposed in \cite{blondel2018nonlinear}, that is the battery model we consider in this work. The hybrid multi-observer is presented in Section~\ref{MultiObserver} and the numerical study based on standard model parameters values is provided in Section~\ref{Simulations}. Finally, Section~\ref{conclusions} concludes the paper. Some details on the model description and the numerical values of the parameters are given in the Appendix. 
\end{comment}

\noindent\textbf{Notation.}
The notation $\R$ stands for the set of real numbers,  $\Rlo:= [0, +\infty)$ and  $\R_{>0}:= (0, +\infty)$. 
We use $\Z$ to denote the set of integers, $\Zo:= \{0,1,2,...\}$ and $\Zp:= \{1,2,...\}$. For a vector $x \in \R^n$, $|x|$ denotes its Euclidean norm. For a matrix $A \in \R^{n  \times m}, \norm{A}$ stands for its 2-induced norm. 
%For a scalar $\xi \in \mathbb{R}$,  $\text{Argsh}(\xi) := \text{ln}(\xi + \sqrt{\xi^2 +1})$.
%
For a signal $v: \R_{\geq0} \to \R^{n_v}$ with $n_v \in \Zp$, and $t_1,t_2\in\R_{\geq 0}\cup\{\infty\}$ with $t_1\leq t_2$,  $\norm{v}_{[t_1, t_2]}:= \textnormal{ess} \sup_{t \in [t_1, t_2]} |v(t)|$. 
Given a set $\mathcal{U} \subseteq \R^{n_u}$, $\mathcal{L}_{\mathcal{U}}$ is the set of all functions from $\R_{\geq0}$ to $\mathcal{U}$ that are Lebesgue measurable and locally essentially bounded. 
Given a real, symmetric matrix $P$, its maximum and minimum eigenvalues are denoted by $\lambda_{\max}(P)$ and $\lambda_{\min}(P)$ respectively. The notation $I_N$ stands for the identity matrix of dimension $N \in \Zp$ and $0_{n\times m}$ stands for the null matrix of dimension $n \times m$, with $n,m \in \Zp$. 
We consider $\K_\infty$ and $\KL$ functions, see \cite[Definitions 3.4 and 3.38]{goebel2012hybrid}.
\section{ELECTROCHEMICAL BATTERY MODEL}\label{BatteryModel}
%In this section we first describe the electrochemical battery model and we state the necessary assumptions used to obtain the state-space form of the considered system. %We then give the state-space form of the system. 
%In this section we first describe the electrochemical battery model together with the assumptions we make to construct it. We then give the model of the considered system in a state-space form.

Before presenting the estimation scheme, we recall the single particle model of lithium-ion battery of \cite{blondel2018nonlinear}.
%We recall in this section the single particle electrochemical model of lithium-ion battery in \cite{blondel2018nonlinear}. %All the parameters, subscripts and indexes are defined in Tables \ref{tab:parameters} and \ref{tab:indexes}.

\begin{figure}[]
	\begin{center}
		\tikzstyle{blockB} = [draw, fill=blue!20, rectangle, 
		minimum height=3.5em, minimum width=7em]  
		\tikzstyle{blockG} = [draw, fill=MyGreen!20, rectangle, 
		minimum height=3.5em, minimum width=8.5em]
		\tikzstyle{blockR} = [draw, fill=red!40, rectangle, 
		minimum height=2em, minimum width=3em]
		\tikzstyle{blockO} = [draw,minimum height=1.5em, fill=orange!20, minimum width=4em]
		\tikzstyle{input} = [coordinate]
		\tikzstyle{blockBattery} = [draw,minimum height=10em, fill=blue!7, minimum width=22em]
		\tikzstyle{blockWplant} = [draw,minimum height=1.5em, fill=white!20, minimum width=4em]
		\tikzstyle{input1} = [coordinate]
		\tikzstyle{blockCircleExternal} = [draw, circle, minimum size=10em]
		\tikzstyle{blockCircleSecond} = [draw, circle, minimum size=8.5em,dashed]
		\tikzstyle{blockCircleThird} = [draw, circle, minimum size=6.5em,dashed]
		\tikzstyle{blockCircleCentral} = [draw, circle, minimum size=2em,dashed]
		\tikzstyle{sum} = [draw, circle, minimum size=.3cm]
		\tikzstyle{blockSensor} = [draw, fill=white!20, draw= blue!80, line width= 0.8mm, minimum height=10em, minimum width=13em]
		\tikzstyle{blockDOT} = [draw,minimum height=8em, fill=white!20, minimum width=14em, dashed]
		\tikzstyle{blockMuxGreen} = [draw, minimum height=21em, fill=MyGreen!30, line width=0.1mm]
		\tikzstyle{blockMuxBlueBig} = [draw, minimum height=21em, fill=blue!30, line width=0.1mm]
		\tikzstyle{blockMuxBlueSmall} = [draw,minimum height=6.5em, fill=blue!30, minimum width=0.8mm]
		\tikzstyle{blockElectrolyte} = [draw,minimum height=1em, fill=blue!7, minimum width=2em]
		
		\begin{tikzpicture}[auto, node distance=2cm,>=latex , scale=0.7,transform shape] 
			
			\node [input, name=input] {};
			\node [blockBattery, right of=input, node distance=14em] (battery){};   
			\node [input, right of=battery, node distance=14em] (output){};   
			\draw [draw,->] (input) -- node [pos=0.4]{\large $u = I$} (battery);
			\draw [draw,->] (battery) -- node [pos=0.6]{\large $y = V$} (output);
			\node [blockElectrolyte, below of=input,  node distance=7.5em,  right =3em] (electrolyte){}; 
			\node [input, right of= electrolyte, node distance=6.50em] (electrolyteName) {};
			\draw [] (electrolyteName) -- node  [pos=0.5]{Electrolyte} (electrolyteName);
			
			\node [input, above of= battery, node distance=5em] (separatorUp) {};
			\node [input, below of= battery, node distance=5em] (separatorDown) {};
			\node [input, left of= separatorUp, node distance=1em] (separatorStartUp) {};
			\node [input, right of= separatorUp, node distance=1em] (separatorEndUp) {};
			\node [input, left of= separatorDown, node distance=1em] (separatorStartDown) {};
			\node [input, right of= separatorDown, node distance=1em] (separatorEndDown) {};
			\draw [draw,-] (separatorStartUp) -- (separatorStartDown);
			\draw [draw,-] (separatorEndUp) -- (separatorEndDown);
			
			\node[blockCircleExternal,left of= battery, node distance=6em, fill=red!20](negElectrodeSurf){};
			\node[blockCircleSecond,left of= battery, node distance=6em](negElectrodeSecond){};
			\node[blockCircleThird,left of= battery, node distance=6em](negElectrodeThird){};
			\node[blockCircleCentral,left of= battery, node distance=6em](negElectrodeCentral){};
			\node[input,above of= negElectrodeSurf, node distance=0.1em](negElectrodeSurfAbove){};
			\path (negElectrodeSurf) -- node[auto=false]{\vdots} (negElectrodeSurfAbove);
			\node [input, above of= negElectrodeCentral, right = 1.2em, node distance=0.0cm] (negElectrodeCentralName) {};
			\draw [] (negElectrodeCentralName) -- node  [pos=0.5]{\textcolor{black}{\textbf{\tiny{$c_1^\text{\tiny neg}$}}}} (negElectrodeCentralName);
			\node [input, above of= negElectrodeCentralName, right = 1.5em, node distance=3.4em] (negElectrodeSecondName) {};
			\draw [] (negElectrodeSecondName) -- node  [pos=0.5]{\textcolor{black}{\textbf{\tiny{$c_{N_{\text{\tiny neg}}-1}^\text{\tiny neg}$}}}} (negElectrodeSecondName);
			\node [input, above of= negElectrodeCentralName, right = 0.6em, node distance=4.4em] (negElectrodeExternalName) {};
			\draw [] (negElectrodeExternalName) -- node  [pos=0.5]{\textcolor{black}{\textbf{\tiny{$c_{N_{\text{\tiny neg}}}^\text{\tiny neg}$}}}} (negElectrodeExternalName);
			\node [input, below of= negElectrodeCentral, right = 4.0em, node distance=6.0em] (negElectrodeName) {};
			\draw [] (negElectrodeName) -- node  [pos=0.5]{Negative electrode} (negElectrodeName);
			
			\node[blockCircleExternal,right of= battery, node distance=6em, fill=MyGreen!20](posElectrodeSurf){};
			\node[blockCircleSecond,right of= battery, node distance=6em](posElectrodeSecond){};
			\node[blockCircleThird,right of= battery, node distance=6em](posElectrodeThird){};
			\node[blockCircleCentral,right of= battery, node distance=6em](posElectrodeCentral){};
			\node[input,above of= posElectrodeSurf, node distance=0.1em](posElectrodeSurfAbove){};
			\path (posElectrodeSurf) -- node[auto=false]{\vdots} (posElectrodeSurfAbove);
			\node [input, above of= posElectrodeCentral, right = 1.2em, node distance=0.0cm] (posElectrodeCentralName) {};
			\draw [] (posElectrodeCentralName) -- node  [pos=0.5]{\textcolor{black}{\textbf{\tiny{$c_1^\text{\tiny pos}$}}}} (posElectrodeCentralName);
			\node [input, above of= posElectrodeCentralName, right = 1.5em, node distance=3.4em] (posElectrodeSecondName) {};
			\draw [] (posElectrodeSecondName) -- node  [pos=0.5]{\textcolor{black}{\textbf{\tiny{$c_{N_{\text{\tiny pos}}-1}^\text{\tiny pos}$}}}} (posElectrodeSecondName);
			\node [input, above of= posElectrodeCentralName, right = 0.6em, node distance=4.4em] (posElectrodeExternalName) {};
			\draw [] (posElectrodeExternalName) -- node  [pos=0.5]{\textcolor{black}{\textbf{\tiny{$c_{N_{\text{\tiny pos}}}^\text{\tiny pos}$}}}} (posElectrodeExternalName);
			\node [input, below of= posElectrodeCentral, right = 4.0em, node distance=6.0em] (posElectrodeName) {};
			\draw [] (posElectrodeName) -- node  [pos=0.5]{Positive electrode} (posElectrodeName);
			
			\node [input, above of= battery, right = -1em, node distance=5.2em] (SeparatorBraceStart) {};
			\node [input, right of= SeparatorBraceStart, node distance=2em] (SeparatorBraceEnd) {};
			\draw [decorate, decoration = {calligraphic brace}] (SeparatorBraceStart)-- (SeparatorBraceEnd); ;
			
			\node [input, above of= SeparatorBraceStart, right = 3.3em, node distance=0.8em] (SeparatorName) {};
			\draw [] (SeparatorName) -- node  [pos=0.5]{Separator} (SeparatorName);
			
			%			\node [input, below of= battery, right = 0.75em, node distance=0em] (SeparatorName) {};
			%			\draw [] (SeparatorName) -- node  [pos=0.5]{$\begin{array}{c}\footnotesize
			%					\text{S} \\[-.1cm]
			%					\text{e} \\[-.1cm]
			%					\text{p} \\[-.1cm]
			%					\text{a} \\[-.1cm]
			%					\text{r}\\[-.1cm]
			%					\text{a}\\[-.1cm]
			%					\text{t}\\[-.1cm]
			%					\text{o}\\[-.1cm]
			%					\text{r}
			%				\end{array} $ } (SeparatorName);
		\end{tikzpicture}
	\end{center}
	\vspace*{-0.6cm} 
	\caption{\normalfont{Battery model schematic}}
	\label{FIG:BatteryModel}
\end{figure}
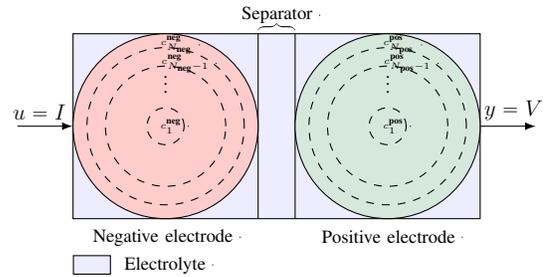

%\subsection{Model description and assumptions} 
%The electrochemical battery model we consider is taken from \cite{blondel2018nonlinear}, which describes the solid diffusion of the lithium that occurs in the solid phase of the electrodes. 
The lithium-ion battery cell, whose schematic is shown in Fig.~\ref{FIG:BatteryModel}, is composed of four elements: two electrodes, one positive and one negative, that are separated by the separator and  %as shown in Figure~\ref{FIG:BatteryModel}. The separator separates the electrodes and 
those three components are immersed in a ionic solution, called electrolyte, which can exchange lithium with the electrodes and provides electrical insulation. Therefore, the electrons cannot be exchanged from one electrode to the other. Due to the electrodes structure, which consists in very small, almost-spherical particles made of porous materials, the electrolyte can penetrate inside the electrode, creating a large contact surface between each electrode and the electrolyte,  which produces an electrochemical coupling between the electrode material and the lithium dissolved in the electrolyte. %is produced. 
Thus, each electrode has a certain potential and this produces
%Thus, there is a large contact surface between each electrode and the electrolyte, which produces a certain potential at each electrode and
a potential difference between the positive and negative electrode. %is produced.  % composed by very small, almost-spherical particles and this 
Since the electrons cannot be exchanged from one electrode to the other within the battery, they will go through an external electrical circuit, if it exists, producing a flow of electrons, that from a macroscopic point of view, corresponds to the current. 
Note that the charges equilibrium in the electrodes and in the electrolyte is preserved at any time because when lithium is removed from its source electrode, another is inserted in its electrode of destination. 
%
% Due to the electrochemical coupling between each electrode and the electrolyte, each electrode has a certain potential.  the electrodes are made of porous materials composed by very small, almost-spherical particles
%
%and consequently there is a potential difference between the positive and negative electrode that produces the electrons flow. 
%
%Each electrode is coupled 
%\begin{figure*}
%	\centering
%	\includegraphics[width=0.6\linewidth]{./Figure/BatteryModel.png}
%	\caption{\normalfont{Battery model schematic \cite{blondel2018nonlinear}
%	}}
%	\label{FIG:BatteryModel}
%\end{figure*}
%
%
%We make the same assumptions as in \cite{blondel2018nonlinear} in order to simplify the electrochemical model and obtain the same state-space form as in \cite{blondel2018nonlinear}. In particular, we assume that the behavior of a single particle of the average size of the particles in the electrode represents the behavior of the whole electrode, as stated in the next assumption.
%
%We consider the lithium-ion battery model given in \cite{blondel2018nonlinear} and consequenly we make the same assumptions. 
%
The model in \cite{blondel2018nonlinear} relies on the next assumption.
%
%\begin{assum}
%	Each electrode in the model is composed of a single sphere particle of the average size of the particles that compose the actual electrode.
%	\hfill $\Box$
%	\label{ASS:singleSphereParticle}
%\end{assum}
%%\textcolor{red}{comment the assumption}
%%
%%To simplify the model, we make also the following assumptions, as in \cite{blondel2018nonlinear}.
%%
%\begin{assum}
%	The electrolyte is neglected. 	\hfill $\Box$
%\end{assum}
%
%\begin{assum}
%	All the parameters of the model are known. 	\hfill $\Box$
%\end{assum}
%
%\begin{assum}
%	The temperature is constant. 	\hfill $\Box$
%\end{assum}

%\begin{ass}
%	The following hold.
%	\begin{enumerate}[label=(\roman*)] %, topsep=0pt,itemsep=-1ex,partopsep=1ex,parsep=1ex]
%		\item 	Each electrode in the model is composed of a single sphere particle of the average size of the particles that compose the actual electrode.
%		\item	The electrolyte dynamics is neglected. 
%		%\item	All the parameters of the model are known. 	\\
%		\item The temperature is constant. \hfill $\Box$
%	\end{enumerate}
%	%
%	\label{ASS:singleSphereParticle}
%\end{ass}

\begin{ass}
	The following hold:
	%\begin{enumerate}[label=(\roman*)] %, topsep=0pt,itemsep=-1ex,partopsep=1ex,parsep=1ex]
			(\textit{i}) %each electrode in the model is composed of a single sphere particle of the average size of the particles that compose the actual electrode;
			the insertion/de-insertion reaction rate is homogeneous throughout the thickness of each electrode; 
		(\textit{ii})	the electrolyte dynamics is neglected;
		%\item	All the parameters of the model are known. 	\\
		(\textit{iii}) the temperature is homogeneous and constant.~\hfill$\Box$
%	\end{enumerate}
	%
	\label{ASS:singleSphereParticle}
\end{ass}
\vspace{-0.6cm}
%\textcolor{red}{Add comments on the assumptions. To design the observer we ignore the electrolyte dynamics, however, in the simulations we consider it.}
Item (\textit{i}) implies that each electrode can be reduced to a single sphere particle of the average size of the particles that compose the actual electrode, which is the single particle model (SPM) as in \cite{blondel2017observer, blondel2018nonlinear,di2010lithium,moura2016battery, dey2015nonlinear}.
%Item (\textit{i}) is the single particle model (SPM) assumption as in \cite{blondel2017observer, blondel2018nonlinear,di2010lithium,moura2016battery, dey2015nonlinear}. 
In view of item~(\textit{ii}) of Assumption~\ref{ASS:singleSphereParticle}, the electrolyte contribution to the output voltage will be represented by a resistive term. However, we will relax this item in the simulation section to evaluate the estimation scheme robustness, see Section~\ref{Simulations}. On the other hand, it is possible to relax the constant temperature assumption in item~(\textit{iii}) of Assumption \ref{ASS:singleSphereParticle} in view of \cite[Sections II.A and III.B]{blondel2018nonlinear}, this is left for future work. 
%For modeling purposes, as in \cite{blondel2018nonlinear}, for the observer design, we ignore the electrolyte dynamics, which is therefore represented with a pure resistance. However, as in \cite{plante2022multiple}, in the simulation in Section~\ref{Simulations}, the electrolyte dynamics is taken into account and thus, the simulation results considers the modeling errors due to this assumption. 
%\textcolor{red}{comment a bit these assumptions. The last one comes from \cite{blondel2017observer} and not \cite{blondel2018nonlinear}, where they relax it... }
%
%\begin{comment}
%\color{blue}

As explained in \cite{blondel2018nonlinear}, in view of item (\textit{i}) of Assumption~\ref{ASS:singleSphereParticle}, the main physical phenomenon is the lithium diffusion in the electrodes, which can be described using partial differential equations \cite{doyle1993modeling, fuller1994simulation}. To simplify the model and obtain a set of ordinary differential equations, each sphere is  spatially discretized in $N_s$ samples of uniform volume, corresponding to $N_s$ crowns, where the subscript ${s} \in \{\textnormal{neg, pos}\}$ denotes the negative or the positive electrode, see Fig.~\ref{FIG:BatteryModel}. We assume for this purpose that the lithium concentration in each crown of the sphere is constant. 
%As in \cite{blondel2018nonlinear}, each sphere is  spatially discretize in $N_s$ samples, where the subscript ${s} \in \{\textnormal{neg, pos}\}$ is introduced for the sake of convenience. We assume that after discretization, the lithium concentration in each crown on the sphere is constant. 
We denote by \(c_i^{\text{s}} \), with \( i \in \{1,\ldots,N_{\text{s}}\} \) and ${s} \in \{\text{neg, pos}\}$, the lithium concentration in the $i^\text{th}$ crown of the electrode, where $i = 1$ corresponds to the one at the center of the electrode, while $i = N_{s}$ corresponds to the one at the surface of the electrode. We also note that the lithium concentration at the center of the negative electrode $c_1^\text{neg}$  can be expressed as a linear combination of all other sampled concentrations in solid phase by mass conservation, and consequently, it does not need to be a system state; see \cite[Section II.C]{blondel2018nonlinear} for more details.
We consider  the state-space model of the lithium-ion battery presented in \cite{blondel2018nonlinear}, where the system state corresponds to the vector of the lithium concentrations in each sample of both electrodes $	x := (c_2^{\text{neg}}, ...,c_{N_{\text{neg}}}^{\text{neg}},c_1^{\text{pos}},...,c_{N_{\text{pos}}}^{\text{pos}})\in \mathbb{R}^{n_x}$, with $n_x := N_{\text{neg}} + N_{\text{pos}} -1$, %\footnote{The lithium concentration at the center of the negative electrode $c_1^\text{neg}$  can be expressed as a linear combination of all other sampled concentrations in solid phase and consequently, it does not need to be a system state.  See \cite[Section II.C]{blondel2018nonlinear} for more details.} 
 the system input $u$ is the current $I$ and the system output $y$ is the output voltage. 
%
%We are ready to present the state-space model of the considered lithium-ion battery. We recall that the system state corresponds to the vector of the lithium concentrations in each sample of both electrodes $	x = (c_2^{\text{neg}}, ...,c_{N_{\text{neg}}}^{\text{neg}},c_1^{\text{pos}},...,c_{N_{\text{pos}}}^{\text{pos}})\in \mathbb{R}^{n_x}  $, the system input $u$ is the current $I$ and the system output $y$ is the output voltage. 
%Let %\begin{equation}
%$	x := (c_2^{\text{neg}}, ...,c_{N_{\text{neg}}}^{\text{neg}},c_1^{\text{pos}},...,c_{N_{\text{pos}}}^{\text{pos}})\in \mathbb{R}^{n_x}  $
%	%\label{eq:x}
%%\end{equation}
%with 
%%\begin{equation}
%	$n_x := N_{\text{neg}}-1 + N_{\text{pos}} $ 
%	%\label{eq:N}
%%\end{equation}
%be the vector of the lithium concentrations in each sample of both electrodes and $u \in \R$ be the input current $I$. Note that, thanks to \eqref{eq:c1}, the dimension of the system state is reduced. 
%\color{blue}
%We are now able to write the following state space equation :
%As in  \cite{blondel2018nonlinear}, 
The model is of the form
\begin{equation} 
	%\  \left\{
	\begin{array}{ll}
		\dot x = Ax+Bu+K + Ev   \\
		y = h(x) + g(u) + w, 
		%
		%	\label{moneq}
	\end{array}
	%\right.
	\label{eq:systeme_state_space}
\end{equation}
%\begin{equation} 
%	%\  \left\{
%	\begin{array}{ll}
%		\dot x = Ax+Bu+K + Ev,   \quad
%		y = h(x) + g(u) + w, 
%		%
%		%	\label{moneq}
%	\end{array}
%	%\right.
%	\label{eq:systeme_state_space}
%\end{equation}
%We refer the reader to \cite{blondel2018nonlinear} for the definitions of the matrices $A\in \mathbb{R}^{n_x\times n_x}$, $B \in \R^{n_x\times 1}$, $K \in \R^{n_x\times 1}$ and for the function $g(u)$ for any $u \in \R$.
The definitions of the matrices $A\in \mathbb{R}^{n_x\times n_x}$, $B \in \R^{n_x\times 1}$, $K \in \R^{n_x\times 1}$ and the function $g(u)$ for any $u \in \R$ are given in Appendix~\ref{appendix_definitions}.
The function $h: \R^{n_x} \to \R$ is defined as, for any $x \in \R^{n_x}$,
%\begin{equation}
$h(x):= \text{OCV}_{\text{pos}}(H_{\text{pos}}x) - \text{OCV}_{\text{neg}}(H_{\text{neg}}x)$, 
%\end{equation} 
with $ H_{\text{neg}} := (0_{ 1\times(N_{\text{neg}}-2)}, \frac{1}{c_{\text{neg}}^{\text{max}}}, 0_{ 1\times N_{\text{pos}}}) \in \mathbb{R}^{1\times n_x}$, $ H_{\text{pos}} := ( 0_{ 1\times (N-1)},$ $\frac{1}{c_{\text{pos}}^{\text{max}}}) \in \mathbb{R}^{1\times n_x}$ and the functions $\text{OCV}_{\text{s}}:\R \to \R$, with $s \in \{\text{neg, pos}\}$ are the open circuit voltages, which are the potential difference between the electrodes and the electrolyte without current and vary with the lithium concentration at the surface of the electrodes. An example of the OCVs is shown in Fig.~\ref{Fig:OCV}. %, which are the functions we consider in the simulations in Section~\ref{Simulations}.
In \eqref{eq:systeme_state_space}, $E \in \R^{n_x \times n_v}$, $v \in \R^{n_v}$ is an unknown disturbance input and $w \in \R$ is an unknown exogenous input affecting the output map. %, which may model mismatch in the current input, as in the simulations in Section~\ref{Simulations}.
%
%where $h: \R^{n_x} \to \R$ is continuous, $v \in \R^{n_v}$ is an unknown disturbance input, which can model errors due to parameter uncertainties, $w \in \R^{n_w}$ is an unknown measurement noise, $E \in \R^{n_x \times n_v}$ and $D \in \R^{1 \times n_w}$. 
%
%
%As in \cite{blondel2018nonlinear}, we obtain
%\begin{equation}
%	\dot x = Ax+Bu+K  
%	\label{eq:state space}
%\end{equation}
%where \(A\in \mathbb{R}^{n_x\times n_x} \) is given in \eqref{eq:A_matrix}, %\(B\in \mathbb{R}^{N} \) and \(K\in \mathbb{R}^{N} \)
%%\begin{equation} 
%%$	B :=\begin{pmatrix}
%%	0_{(N_{Neg-2})\times 1} &
%%	- \bar{K}_I^{\text{neg}} &
%%	0_{(N_{pos-1})\times 1}&
%%	\bar{K}_I^{\text{pos}} &
%%\end{pmatrix}^\top \in \R^{n_x\times 1}$ 
%$	B :=(
%0_{(N_{Neg-2})\times 1} \; 
%- \bar{K}_I^{\text{neg}} \;
%0_{(N_{pos-1})\times 1} \;
%\bar{K}_I^{\text{pos}} \;
%)^\top \in \R^{n_x\times 1}$
%and
%%	\label{eq:B}
%%\end{equation}
%%\begin{equation} 
%%$	K :=\begin{pmatrix}
%%	-	k_1^{\text{neg}}\tau_2^{\text{neg}}\bar{K}&
%%	0_{(N-1)\times 1} 
%%\end{pmatrix}^\top \in \R^{n_x\times 1}$,
%$	K :=(
%-	k_1^{\text{neg}}\tau_2^{\text{neg}}\bar{K} \;$ $
%0_{(N-1)\times 1} 
%)^\top \in \R^{n_x\times 1}$,
%%\label{eq:K}
%%\end{equation}
%$h: \R^{n_x} \to \R$ is defined as $h(x):= OCV_{\text{pos}}(H_{\text{pos}}x) - OCV_{\text{neg}}(H_{\text{neg}}x)$,
% $v \in \R^{n_v}$ is an unknown disturbance input, which can model errors due to parameter uncertainties, $w \in \R^{n_w}$ is an unknown measurement noise, $E \in \R^{n_x \times n_v}$ and $D \in \R^{1 \times n_w}$.
  We assume that $u: \R_{\geq 0} \rightarrow \R^{n_u}$,  $v:\R_{\geq 0} \rightarrow \R^{n_v}$ and  $w:\R_{\geq 0} \rightarrow \R$ are such that $u \in \mathcal{L_U}$, $v \in \mathcal{L_V}$ and $w \in \mathcal{L_W}$ for closed sets $\mathcal{U} \subseteq \R^{n_u}$, $\mathcal{V} \subseteq \R^{n_v}$ and $\mathcal{W} \subseteq \R$, which is very reasonable for lithium-ion batteries.
The lithium concentrations in the electrodes are related to the state of charge (SOC) of the battery, which is an essential information. Indeed, the SOC is defined as, for all $t \geq 0$, 
\begin{equation}
	\text{SOC}(t) :=   100 \frac{\bar{c}^{\text{pos}}(t)-c_0^{\text{pos}}}{c_{100}^{\text{pos}}-c_0^{\text{pos}}}
	\label{eq:SOC}
\end{equation}
with $\bar{c}^{\text{pos}}(t) := \frac{1}{V_{\text{total}}^{\text{pos}}}\sum_{i=1}^{N_{\text{pos}}} c_i^{\text{pos}}(t) V_i^{\text{pos}}$, where $c_0^\text{pos}$ and $c_{100}^\text{pos}$ are the lithium concentrations in the positive electrode at $\text{SOC} = 0 \ \%$ and at $\text{SOC} = 100 \ \%$, respectively,  $V_{\text{total}}^{\text{pos}}$ is the total volume of the positive electrode and $V_{\text{i}}^{\text{pos}}$ is the volume of the $i^\text{th}$ sample of the positive electrode. %Consequently, designing an observer to estimate the lithium concentrations of the battery, which cannot be measured directly, is important to obtain an estimation of the state of charge of the battery. 
The concentrations in the positive electrode are considered in \eqref{eq:SOC}; the same value for the SOC would be obtained by considering the concentrations in the negative electrode. Hence by estimating the concentrations in the electrodes, we will be able to also estimate the SOC.
%\color{black}
%
We now design an estimation scheme for this purpose.

%\textcolor{blue}{Add a transition between the two sections, like: only u and y are known, while we want to know the concentrations $->$ we design an estimation scheme for this purpose.}

\section{HYBRID MULTI-OBSERVER DESIGN}\label{HybridMultiObserver}
%\color{blue}
%To estimate the concentrations in the electrodes of a lithium-ion battery model \eqref{eq:systeme_state_space},  an observer, based on a polytopic approach, was designed in \cite{blondel2017observer}. However, it is very difficult to tune its performance in terms of convergence speed and robustness to measurement noise $w$ and disturbance $v$. To address this challenge, we apply the scheme we recently presented in \cite{petri2022towards}, which exploits hybrid techniques and a multi-observer to improve the estimation performance. 
The hybrid multi-observer consists of the following elements:
\begin{itemize}
	\item \textit{nominal observer}, here we consider the one proposed in \cite{blondel2017observer}, which satisfies an input-to-state stability property, as required by \cite[Assumption 1]{petri2022towards}; 
	\item $N$ additional dynamical systems with the same structure as the nominal observer, but with a different output injection gain, that can be arbitrarily selected. % having an observer structure. Note that they are not required to guarantee a convergence property of the estimation error and, thus, each of them can be an asymptotic observer or not.
	%These systems, together with the nominal observer, form the \textit{multi-observer}, where each element is called \textit{mode} for the sake of convenience;
	Each of these systems, as well as the nominal observer, is called \textit{mode} for the sake of convenience;
	\item \textit{monitoring variables} used to evaluate the performance of each mode of the multi-observer;
	\item \textit{selection criterion}, that selects one mode of the multi-observer at any time instant, based on the performance knowledge given by the monitoring variables;  %switches between the state estimates produced by the different modes exploiting the performance knowledge given by the monitoring variables;
	\item \textit{reset rule}, that updates the estimation scheme when a switching of the selected mode occurs; 
	\item \textit{filtered version} of the hybrid multi-observer state estimate to produce a continuous state estimate. %, that solves the possible issue of discontinuity. 
	Note that this is a novelty compared to~\cite{petri2022towards}. 
\end{itemize}

\subsection{Nominal observer}\label{NominalObserver}
%From \eqref{eq:state space} and \eqref{eq:y}, the model of lithium-ion battery is of the form
%\begin{equation} 
%	%\  \left\{
%	\begin{array}{ll}
%		\dot x = Ax+Bu+K+Ev    \\
%		y = h(x) + g(u) + Dw,
%		%
%	%	\label{moneq}
%	\end{array}
%	%\right.
%	\label{eq:systeme_state_space}
%\end{equation}
%where $h: \R^{n_x} \to \R$ is continuous, $v \in \R^{n_v}$ is an unknown disturbance input, which can model errors due to parameter uncertainties, $w \in \R^{n_w}$ is an unknown measurement noise, $E \in \R^{n_x \times n_v}$ and $D \in \R^{1 \times n_w}$. 

Inspired by \cite{blondel2017observer}, we design a nominal observer that satisfies the input-to-state stability property in \cite[Assumption 1]{petri2022towards}. We make the next assumption for this purpose.
\begin{ass}
	The parameters of the model are known.\hfill$\Box$
\end{ass}
The nominal observer has the form %to estimate the lithium concentrations of the battery
\begin{equation} 
	%	\  \left\{
	\begin{array}{ll}
		\dot{\hat{x}}_1 = A\hat{x}_1+Bu+K+L_1(y-\hat{y}_1)  \\
		\hat{y}_1 = h(\hat{x}_1) + g(u), 
		%	\label{moneq}
	\end{array}
	%	\right.
	\label{eq:systeme_obs}
\end{equation}
where $\hat{x}_1 \in \R^{n_x}$ is the state estimate, %\footnote{The subscript $1$ is used because the nominal observer in \eqref{eq:systeme_obs} is the first element of the multi-observer that we will design in Section~\ref{MultiObserver}.},
$\hat{y}_1\in \R$ is the estimated output and $L_1 \in \R^{n_x \times 1}$ is the output injection gain, that needs to be designed; we use the subscript $1$ because the nominal observer in \eqref{eq:systeme_obs} is the first element of the multi-observer that we will design in Section~\ref{MultiObserver}.
%\begin{rem}
% The state estimate of the nominal observer \eqref{eq:systeme_obs} has the same dimension $n_x$ as the system state \eqref{eq:systeme_state_space}. This is because the observer is designed based on the system model, which is obtained discretizing each sphere of the electrodes to obtain a set of ordinary differential equations. Consequently, the higher is the number of samples, the closer is the model to the real system. To introduce some model uncertainties in the simulations, to evaluate the efficiency of the observer, we can consider a 
%\hfill $\Box$
%\label{rem:modelUncertaintesRemark}
%\end{rem}
%
%Note that we use the subscript $1$ because the nominal observer \eqref{eq:systeme_obs} is the first element of the multi-observer that we will design in Section~\ref{MultiObserver}.
While \eqref{eq:systeme_obs} involves the plant input $u$, possible mismatches on the input current known by the plant and the observer, which often occur in practice, can be modeled using the disturbance input $v$ and the exogenous input $w$ in \eqref{eq:systeme_state_space}, as we will do in Section~\ref{Simulations}. 
We define the state estimation error as $e_1:=x-\hat{x}_1$. As in \cite[Assumption~1]{petri2022towards}, we define a perturbed version of the $e_1$-dynamics, which is given by, in view of \eqref{eq:systeme_state_space} and \eqref{eq:systeme_obs}, 
\begin{equation}
	\dot{e}_1 = Ae_1 +Ev -L_1(h(x)-h(\hat{x}_1))-L_1w + d,
	\label{eq:erreur}
\end{equation}
%\color{blue} 
where $d \in \R^{n_x}$ represents an additional artificial perturbation on the output injection term $L_1(y - \hat y_1)$. %, that can be used to model possible mismatch in the current input, as we will do in the simulations in Section~\ref{Simulations}. 
To consider the perturbed dynamics in \eqref{eq:erreur} with extra input $d$ is required to check one of the key assumptions of \cite{petri2022towards}, which is needed to establish the main result of the work. % allowed to prove an input/output-to-state stability property of the additional modes \cite{sontag2008input}}.

We design the observer gain $L_1$ to guarantee a convergence property of the estimation error $e_1$. In particular, $L_1$ has to be designed such that the origin of~\eqref{eq:erreur} satisfies an input-to-state stability property with respect to $v$, $w$ and $d$. To design the observer gain $L_1$, we make the next assumption on the OCVs, which is taken from \cite[Assumption 5]{blondel2017observer}.
\begin{ass}
There exist constant matrices $C_1, \dots, C_{4} \in \R^{1 \times n_x}$ such that, for any $x$, $x' \in \R^{n_x}$,	%For any $x$, $x' \in \R^{n_x}$ there exist constant matrices $C_1, \dots, C_{4} \in \R^{1 \times n_x}$ such that 
	\begin{equation}
		h(x) -h(x') = %OCV_{\text{pos}}(H_{\text{pos}}x)-OCV_{\text{neg}}(H_{\text{neg}}x) - 	OCV_{\text{pos}}(H_{\text{pos}}x)+OCV_{\text{neg}}(H_{\text{neg}}x) = 
		C(x,x')(x-x'), 
	\end{equation}
	where $C(x,x'):= \sum_{i = 1}^{4}\lambda_i(x,x')C_i$, with $\lambda_i \in [0,1]$, $\sum_{i = 1}^{4}\lambda_i(x,x') =1$ and $i \in  \{1, \dots, 4\}$.
	\hfill $\Box$
	\label{ASS:OCVassumption}
\end{ass}
Assumption~\ref{ASS:OCVassumption} means that the output map $h$ lies in a polytope defined by the vertices $C_i$, with $i \in \{1,\dots, 4\}$. This condition is often verified in practice. Indeed, the OCVs are generally defined on the interval $[0,1]$ by experimental data and they are well-approximated by a piecewise continuously differentiable function. Moreover, the OCVs only depend on the surface lithium concentration of the negative and positive electrode. Consequently, the output map $h$ only depends on two states of the system and the set of $C_i$ has only $2^2$ elements, which are obtained from the maximum and minimum slopes of the OCVs. 
%\color{black}
%Note that the output map $h$ of system \eqref{eq:systeme_state_space} satisfies \cite[Assumption 5]{blondel2018nonlinear}. Indeed $h$ is continuously differentiable and $h(x) = \sum_{i=1}^{n_x}h_i(x_i)$, with $\underline{h_i} \leq \frac{\partial h_i(x_i)}{\partial x_i} \leq \overline{h_i}$ for almost all $x_i \in \R$ with $\underline{h_i}, \overline{h_i} \in \R$. 
%
Using Assumption~\ref{ASS:OCVassumption}, \eqref{eq:erreur} becomes
%\textcolor{red}{Do we need to write and explain \cite[Assumption 5]{blondel2018nonlinear}??. %Why $n_x$ if then it is not $n_x$ in the example??
%}
%
\begin{equation}
	\dot{e}_1 = (A-L_1C(x,\hat{x}_1))e + Ev-L_1w -d.
	\label{eq:erreur_avec_hypothese}
\end{equation}
To design the observer output injection gain $L_1$ we follow a polytopic approach and we propose a modified version of \cite[Theorem 1]{blondel2017observer} below.  
\begin{thm}
	Consider system \eqref{eq:erreur_avec_hypothese}. % and let $\mathcal{H}_i:= (A-L_1C_i)^\top P + P(A-L_1C_i)$, with $i \in \{1, \dots, 4\}$.
	 If there exist $L_1 \in \R^{n_x \times 1}$, $\alpha$,  $\mu_v$, $\mu_w \text{ and } \mu_d \in \R_{>0}$ and $P \in \R^{n_x \times n_x}$ symmetric positive definite such that %for all $i \in \{1, \dots,4\}$
	\begin{equation}
%		\begin{pmatrix}
%			\mathcal{H}_i + \alpha P  & - PL_1D & -P\\
%			-D^\top L_1^\top P  & -\mu_w I_{n_w} & 0\\
%			-P  & 0 & -\mu_d I_{n_x}
%		\end{pmatrix} \leq 0,
\small
	\begin{pmatrix}
		\mathcal{H}_i + \alpha P & PE & - PL_1 & -P\\
		E^\top P & -\mu_v I_{n_v} &0 & 0\\
		- L_1^\top P & 0 & -\mu_w I_{n_w} & 0\\
		-P & 0 & 0 & -\mu_d I_{n_x}
	\end{pmatrix} \leq 0,
	\label{eq:LMI_ThISSNominal}
\end{equation}
	with $\mathcal{H}_i:= (A-L_1C_i)^\top P + P(A-L_1C_i)$ for all $i \in \{1, \dots, 4\}$. Then $V: e_1 \mapsto e_1^\top P e_1$ satisfies, for any $e_1 \in \R^{n_x}$, $v \in \mathcal{V}$, $w \in \mathcal{W}$ and $d \in \R^{n_x}$, 
	\begin{equation}
		\lambda_{\text{min}}(P) |e_1|^2 \leq V(e_1) \leq \lambda_{\text{max}}(P)|e_1|^2,
		\label{eq:nominalObserverSandwichBound}
	\end{equation}
	\begin{equation}
		\begin{array}{l}
			%	\hspace{-0.5em} 
			\left\langle \nabla V(e_{1}), (A-L_1C(x,\hat x_1))e_1  - L_1w \right\rangle  \\
		\qquad \quad	\leq 
			-\alpha V(e_{1})  + \mu_v|v|^2+ \mu_w|w|^2 + \mu_d|d|^2. 
		\end{array}
		\label{eq:ISS_nominalObserver}
	\end{equation}
\hfill$\Box$
	\label{thm:observerTheorem}
\end{thm}
The proof of Theorem~\ref{thm:observerTheorem} follows similar steps as \cite[proof of Theorem 1]{blondel2017observer} and is therefore omitted. 
%\textcolor{red}{Since the proof is very similar to the one in that paper, I think it doesn't make sense write it here, but, in case you want to read it, I wrote it in the appendix}. 
Theorem~\ref{thm:observerTheorem} guarantees that the nominal observer \eqref{eq:systeme_obs} satisfies an input-to-state stability property with respect to the disturbance $v$, the exogenous perturbation $w$ and the additional perturbation $d$. % and therefore, its estimation error converges. . 
This implies that the estimation error $e_1$ exponentially converges to a neighborhood of the origin, whose ``size'' depends on the $\mathcal{L}_\infty$ norm of $v$, $w$ and $d$.
As a result, \cite[Assumption~1]{petri2022towards} is satisfied. 
%Note that, \cite[Assumption~1]{petri2022towards} consider a perturbed version of the error dynamic and, consequently, the input-to-state stability property is required with respect also to an additional perturbation $d$, which, however, acts as an additive measurement noise. Thus, \cite[Assumption~1]{petri2022towards} is satisfied by Theorem~\ref{thm:observerTheorem} also when a perturbed version of the error dynamics is considered. 
%\color{blue}%MODIFY THIS PART ACCORDING TO ROMAIN'S COMMENT
%Moreover, from Assumption \ref{ASS:OCVassumption} and considering the quadratic Lyapunov function $V$ from Theorem~\ref{thm:observerTheorem}, \cite[Assumption~2]{petri2022towards} is satisfied. We can now design the hybrid multi-observer.
%However, it is very difficult to tune
%
The possible drawback of observer \eqref{eq:systeme_obs} with $L_1$ designed as in Theorem~\ref{thm:observerTheorem} is that to tune the observer gain $L_1$ to obtain good estimation performance both in speed of convergence and robustness to measurement noise, exogenous perturbation and disturbance is very difficult in general. For this reason, we apply our recent result in~\cite{petri2022towards}, which consists in designing a hybrid multi-observer with the aim of improving the estimation performance of \eqref{eq:systeme_obs}.
%\color{black}

%\color{blue}
%Theorem 1 in \cite{blondel2017observer} guarantees that, if the linear matrix inequality in \cite[Equation (16)]{blondel2017observer} is satisfied, then system \eqref{eq:erreur_avec_hypothese} is $\mathcal{L}_2$-stable from $v$ and $w$ to $e_1$, i.e., there exists $c\geq 0$ such that, for any initial condition $e_{1_0} \in \R^{n_x}$ and any $v, w \in \mathcal{L}_2$, the corresponding solution $e_1$ to \eqref{eq:erreur_avec_hypothese} verifies, for any $t \geq 0$, 
%\begin{equation}
%	\norm{e_1}_{2[0,t)} \leq c|e_{1_0}| + \sqrt{\frac{\mu_v}{\varepsilon}}\norm{v}_{2[0,t)} + \sqrt{\frac{\mu_w}{\varepsilon}}\norm{w}_{2[0,t)},
%%	\label{eq:ISS_nominalObserver}
%\end{equation} 
%for some $\mu_v, \mu_w, \varepsilon \in \R_{>0}$. 
%
%%Note that, \cite[Theorem 1]{blondel2017observer} requires solving a linear matrix inequality, and consequently, only if it has a solution, \cite[Theorem 1]{blondel2017observer} guarantees the stability property in \eqref{eq:ISS_nominalObserver}. However, we are considering the same system model as in \cite{blondel2017observer} and in this case the linear matrix inequality in  \cite[Theorem 1]{blondel2017observer} is feasible and thus we obtain the observer gain $L$.
%\textcolor{red}{Probably we don't need to rewrite what they guarantee, but we can just write that, inspired by \cite[Theorem 1]{blondel2017observer} we design the gain $L_1$ with the modified LMI (which is not anymore an LMI) and we guarantee an ISS property.}
%\color{black}

\subsection{Hybrid multi-observer}\label{MultiObserver}
To improve the estimation performance of the nominal observer \eqref{eq:systeme_obs}, we design the  hybrid multi-observer proposed in \cite{petri2022towards}. For this purpose, we consider $N$ additional dynamical systems with the form of \eqref{eq:systeme_obs}, where the number $N \in \Z_{>0}$ is freely selected by the user, but with a different output injection gain, i.e., for any $k \in \{2, \dots, N+1\}$, the $k^{\text{th}}$ mode of the multi-observer is given by
\begin{equation} 
	%	\  \left\{
	\begin{array}{ll}
		\dot{\hat{x}}_k = A\hat{x}_k+Bu+K+L_k(y-\hat{y}_k)  \\
		\hat{y}_k = h(\hat{x}_k) + g(u), 
		%	\label{moneq}
	\end{array}
	%	\right.
	\label{eq:ObserverAdditional}
\end{equation}
where $\hat{x}_k \in \R^{n_x}$ is the mode $k$ state estimate, $\hat{y}_k \in \R$ is the mode $k$ estimated output and $L_k \in \R^{n_x \times 1}$ is its output injection gain. %We define their state estimation errors as $e_k:= x-\hat{x}_k$, with $k \in \{2, \dots, N+1\}$. 
Since there is full freedom on the selection of the gains $L_k$, with $k \in \{2, \dots, N+1\}$, there are no convergence guarantees on the estimation errors $e_k:= x-\hat{x}_k$, with $k \in \{2, \dots, N+1\}$.  %We believe that this freedom is an interesting feature of the hybrid multi-observer. Indeed, the gains obtained using most of the observer design techniques in the literature that guarantee a convergence property of the estimation error are typically very conservative and this affects the performance. Therefore, selecting the gains without caring about their convergence property allows to be less conservative and produces different behaviors, that may help to address the different trade-offs of the state estimation problem. 
%Using \eqref{eq:LMI_ThISSNominal}, it is difficult to tune the nominal observer gain $L_1$ to obtain good estimation performance. Consequently, it is difficult also to select the gains $L_k$, with $k \in \{2, \dots, N+1\}$ of the additional modes. However, we can exploit the freedom we have on their selection and we can choose the additional gains considering the behaviour of the nominal observer. 
	%
%	To select the gains $L_k$, with $k \in \{2, \dots, N+1\}$ we can look at the behaviour of the nominal observer.
%	Indeed, in an observer design typically a large gain produces a fast convergence, but high sensitivity to noises and a small gain typically is more robust to noise, but generates a slow convergence. 
%	\textcolor{blue}{As mention above, the gains of the additional modes can be freely selected. As a consequence, it is possible to select them randomly. However, following some criterion for their selection should allow to obtain better results. }
	A recommended approach to select the gains  $L_k$'s is to  consider the behaviour of the nominal observer in \eqref{eq:systeme_obs} in simulation and, based on that, to select the additional gains depending on the property we want to improve. 
	For instance, if the convergence speed of the estimation error $e_1$ is too slow, we may define the $L_k$ by increasing the values of $L_1$. On the opposite, if the convergence speed of $e_1$ is satisfactory but its accuracy for large time is not satisfactory, we may select the gains $L_k$ with small values, as we will do in Section~\ref{Simulations}.
	There are many other approaches that can be followed to select the additional gains. For example, we may pick them in a neighborhood of the nominal one or design one additional gain for each vertex of the polytope.
	Note that these gain selection criteria may result in diverging estimation errors for some of the modes, still the overall hybrid scheme we present does ensure the (approximate) convergence of the obtained state estimation error to the origin.
	%\textcolor{red}{Aggiungi qualche commento su altre tecniche per la scelta dei gain... vedi Rev 3, commento 5 e prendi spunto dal TAC. }
	%However, this is not an issue in view of the freedom we have from \cite{petri2022towards}. In the numerical study in Section~\ref{Simulations} we select the additional gains using this approach. 
%Moreover, 
%However, \cite[Theorem~1]{petri2022towards} gives us the guarantee that the obtained hybrid multi-observer satisfies an input-to-state stability property, and thus that its estimation error converges.  

To select which state estimate $\hat x_k$, $k\in \{1, \dots N+1\}$, we need to consider, we evaluate which mode has the best performance. To define performance, we introduce monitoring variables,
%The hybrid multi-observer design requires the use of monitoring variables, 
denoted  $\eta_k \in \R_{\geq 0}$, with $k\in \{1, \dots, N+1\}$, %, to evaluate the performance of each mode. 
whose dynamics are
\begin{equation}
	\begin{aligned}
		\dot{\eta}_k &= -\nu\eta_k + \lambda_1|y-\hat{y}_k|^2 + \lambda_2|L_k(y-\hat{y}_k)|^2,%\\ 
		%	&=: g(\eta_k, L_k, y, y_k ), %\quad %\eta_k(0) = 0,
	\end{aligned}
	\label{eq:etaDynamics}
\end{equation}
where $\lambda_1, \lambda_2 \in \R_{\geq 0}$, with $\max(\lambda_1,\lambda_2) >0$ and $\nu \in (0,\alpha]$, are design parameters, with $\alpha$ from Theorem~\ref{thm:observerTheorem}. The condition $\nu \in (0,\alpha]$ is required to establish the stability property formalized in \cite{petri2022towards}. The monitoring variable dynamics is inspired by \cite{willems2004deterministic} and depends both on output estimation error, with the term $\lambda_1 |y-\hat y_k|^2$, and on the correction effort of the observer, with $\lambda_2 |L_k(y-\hat y_k)|^2$. This last term is called latency in \cite{willems2004deterministic}. Equation~\eqref{eq:etaDynamics} implies that, for any $k \in \{1, \dots, N+1\}$, for any initial condition $\eta_k(0) \in \R_{\geq 0}$, for any $y, \hat{y}_k \in \mathcal{L}_{\R^{n_y}}$ and any $t \geq 0$, 
\begin{equation}
	\begin{aligned}
		\eta_k(t) = & \ e^{-\nu t}\eta_k(0) + 
		\int_{0}^{t}e^{-\nu(t-\tau)}\left( \lambda_1|y(\tau)-\hat{y}_k(\tau)|^2 \right. \\
		&+  \left. \lambda_2|L_k(y(\tau)-\hat{y}_k(\tau))|^2  \right)d\tau.
	\end{aligned}
	\label{eq:etaDynamicsTime}
\end{equation}
%\textcolor{red}{Write some comments on the fact that we can tune the initial condition of $\eta_k$.}
From \eqref{eq:etaDynamicsTime} we have that the monitoring variables represent the cost of the modes. Consequently, the idea is to select the mode that produces the minimum monitoring variable, and thus the minimum cost, at any time instant. % that minimize its monitoring variable, in order to minimize such a  cost. 
%\color{blue}
Note that, we can freely choose the initial conditions of these monitoring variables, $\eta_k(0) \in \R_{\geq 0}$, with $k \in \{1, \dots, N+1\}$. This extra degree of freedom can be used to initially select or penalize one or more modes of the multi-observer, as we do in simulation in Section~\ref{Simulations}.  
%\color{black}
The signal $\sigma:\R_{\geq 0} \to \{1, \dots, N+1\}$ is used to indicate the selected mode at any time instant. The corresponding state estimate, monitoring variable and state estimation error are denoted $\hat{x}_\sigma$,  $\eta_\sigma$ and $e_\sigma$, respectively. We denote with $t_0 = 0$ the initial time and with $t_i \in \R_{\geq 0}$, $i \in \Zp$ the times when a switch of the selected mode occurs, i.e.,  $t_{i}:= \inf \{t \geq t_{i-1}: \exists k \in \{1,\dots, N+1\} \setminus \{\sigma(t)\}  \text{ such that } \eta_k(t) \leq \varepsilon \eta_{\sigma(t)}(t)\}$, where $\varepsilon \in (0,1]$ is a design parameter introduced to mitigate the occurrence of fast switching. Consequently, for all $i\in \Zp$, $\dot \sigma(t) = 0$ for all $t \in (t_{i-1}, t_{i})$ and $\displaystyle \sigma(t_{i}) \in \operatornamewithlimits{arg\,min}\limits_{k \in \{1, \dots, N+1\}} \eta_k(t_{i})$.
%
%\color{blue}
Finally, when switching occur, %as proposed in \cite{petri2022towards}\footnote{In \cite{petri2022towards} the authors proposed two possible reset rules, called without and with resets. Here we consider the reset option.}, 
not only the signal $\sigma$ is updated, but also the state estimates and the monitoring variables of the additional modes are reset to the ones of the selected mode, \footnote{In \cite{petri2022towards} two possible reset rules are considered, called without and with resets. Only the reset strategy is considered in this work.}, 
i.e., at a switching time $t_i \in \R_{\geq 0}$, $i \in \Zp$,
\begin{equation}
	\begin{aligned}
		&\hat x_k(t_i^+) \in \{\hat x_{k^\star}(t_i): 
		k^\star \in \!\!\! \operatornamewithlimits{arg\,min}\limits_{j \in \{ 1, \dots, N +1\}} \eta_j(t_i) \},
	\end{aligned}
	\label{eq:StateEstimatek_plus_reset}
\end{equation}
\begin{equation}
	\begin{aligned}
		&\eta_k(t_i^+) \in \{\eta_{k^\star}(t_i): 
		k^\star \in  \!\!\! \operatornamewithlimits{arg\,min}\limits_{j \in \{ 1, \dots, N +1\}} \eta_j(t_i) \},
	\end{aligned}
	\label{eq:Eta_plus_reset}
\end{equation}
where $t_i^+$ represent the time immediately after the switching instant $t_i$ and $k \in \{2, \dots, N+1\}$.
The state estimate $\hat{x}_\sigma$ produced by the hybrid multi-observer may be discontinuous. %at switching times, which may be a problem for batteries as this means the SOC estimate may experience jump. Indeed, having the state of charge estimate that jumps can confuse the user.
 For this reason, we add a filtered version of $\hat{x}_\sigma$, denoted $\hat{x}_f$, whose dynamics between switching is given by
\begin{equation}
	\dot{\hat{x}}_f = - \zeta\hat{x}_f + \zeta\hat{x}_\sigma,
	\label{eq:filteredStateEstimateFlow}
\end{equation}
where $\zeta > 0$ is an additional design parameter and, at switching times $t_i \in \R_{> 0}$, with $i \in \Zp$,
\begin{equation}
	\hat{x}_f(t_i^+) = \hat{x}_f(t_i).
	\label{eq:filteredStateEstimateJump}
\end{equation}

\subsection{Hybrid model and stability guarantees}
Including $\hat{x}_f$, we obtain a new hybrid model for the hybrid multi-observer compared to \cite{petri2022towards}, whose state is defined as ${q}:=  (x,\hat{x}_1, \dots, \hat{x}_{N+1}, \eta_1, \dots, \eta_{N+1}, \sigma, \hat{x}_f)\in {\mathcal{Q}}:= \R^{n_x} \times \R^{(N+1)n_x} \times \R_{\geq 0}^{N+1}  \times \{1, \dots, N +1\} \times \R^{n_x}$. The hybrid system is given by
	\begin{equation}
	%\mathcal{H}(F,G,\mathcal{C},\mathcal{D}) : \ 
	\left\lbrace \
	\begin{aligned}
		\dot{{q}} &=  F({q}, u,v, w), \ \ \ \ \ && q \in {\mathcal{C}}\\
		{q}^{+} &\in  G( q), \ \ \ \ \ &&  q \in {\mathcal{D}},
	\end{aligned}
	\right.
	\label{eq:HybridSystemGeneral}
\end{equation}
where the flow map $ F$ is obtained from \eqref{eq:systeme_state_space}, \eqref{eq:systeme_obs}, \eqref{eq:ObserverAdditional}, \eqref{eq:etaDynamics} and \eqref{eq:filteredStateEstimateFlow}, the jump map  $ G$ follows from the above developments, \eqref{eq:filteredStateEstimateJump} and is similar to the jump map in \cite{petri2022towards}.  %is obtained from \eqref{eq:filteredStateEstimateJump} and the jump map in \cite{petri2022towards}, 
The flow and jump sets, ${\mathcal{C}}$ and ${\mathcal{D}}$, are defined as
\begin{align}
\mathcal{C} &:= \left\{q \in \mathcal{Q}: \forall k \in \{1, \dots, N+1\} \:  \ \eta_k \geq  \varepsilon\eta_\sigma \right\}\!,	\label{eq:flowSet} \\
\mathcal{D} &:= \left\{q \in \mathcal{Q}: \exists k \in \{1, \dots, N+1\} \setminus \{\sigma\} \: \ \eta_k \leq \varepsilon\eta_\sigma \right\}\!.
	\label{eq:FlowAndJumpSets}
\end{align}
%contain the same condition as $\mathcal{C}$ and $\mathcal{D}$ in \cite{petri2022towards} and the only difference is that they are defined for the new hybrid state $\tilde q$. \textcolor{red}{I tried to be concise and not write the definitions because probably the paper is already too long, but I'm not sure if it is enough clear the hybrid system defined like that. What do you think? }

%In the next theorem, which is an extension of \cite[Theorem 1]{petri2022towards}, we prove that system \eqref{eq:HybridSystemGeneral} satisfies a two-measure input-to-state stability property with respect to the disturbance input $v$ and measurement noise $w$ \cite{cai2007smooth}. 
The next theorem ensures that system \eqref{eq:HybridSystemGeneral} satisfies a two-measure input-to-state stability property with respect to the disturbance $v$ and the perturbation $w$ \cite{cai2007smooth}. %\textcolor{red}{Its proof follows similar lines as \cite[Proof of Theorem 1]{petri2022towards} and is therefore omitted.}
	\begin{thm}
	Consider system \eqref{eq:HybridSystemGeneral} and suppose Assumptions~\ref{ASS:singleSphereParticle}-\ref{ASS:OCVassumption} hold and $L_1$ is selected such that
	condition~\eqref{eq:LMI_ThISSNominal} in Theorem~\ref{thm:observerTheorem} is satisfied. % and Assumptions~\ref{ASS:singleSphereParticle}-\ref{ASS:OCVassumption} hold. %Select $c_1, c_2, c_3, a, b$ like in Theorem~\ref{THM:LyapunovTheorem}. 
	Then, there exist $\beta_U \in \KL$ and $\gamma_U \in \Kinf$ such that for any input $u \in \mathcal{L_U}$, disturbance input $v \in \mathcal{L_V}$ and exogenous perturbation $w \in \mathcal{L_W}$, any solution $q$ satisfies %for all $(t,j) \in \dom q$, 
	\begin{equation}
		\begin{array}{l}
			|(e_1(t,j), \eta_1(t,j), e_{\sigma} (t,j), \eta_{\sigma}(t,j), e_f(t,j))|  \\[0.5em]
			\qquad \leq \beta_U(|(e(0,0), \eta(0,0))|,t) + \gamma_U(\norm{v}_{[0,t]} + \norm{w}_{[0,t]})
		\end{array}
		\label{eq:LyapunovSolutionTheoremEquation}
	\end{equation}
	for all $(t,j)$ in the domain\footnote{The solution $q$ of the hybrid multi-observer is defined on hybrid-time domains, see \cite[Definition 2.3]{goebel2012hybrid}, where the first argument is the continuous time $t$, while the second argument is the discrete time $j$ and represents the number of jumps/switching.} of the solution $q$, with $e:= (e_1, \dots, e_{N+1})$, $\eta:= (\eta_1, \dots, \eta_{N+1})$, $e_\sigma := x- \hat{x}_\sigma$ and $e_f := x-\hat{x}_f$.
	\hfill $\Box$
	%\textcolor{red}{I should define $e:= (e_1, \dots e_{N+1})$ and $\eta:= (\eta_1,\dots, \eta_{N+1})$ and use this here and in the proof to have a more compact notation.}
	\label{thm:LyapunovSolutionProposition}
\end{thm}
\noindent\textbf{Sketch of proof:}
%\color{blue}
We first note that all the conditions of \cite[Theorem 1]{petri2022towards} are satisfied. Indeed, thanks to Theorem~\ref{thm:observerTheorem}, \cite[Assumption 1]{petri2022towards} holds. Moreover, \cite[Assumption 2]{petri2022towards} is satisfied thanks to Assumption~\ref{ASS:OCVassumption} and because the Lyapunov function $V$ in Theorem~\ref{thm:observerTheorem} is quadratic. We can then follow similar steps as in \cite[proofs of Proposition 1 and Theorem 1]{petri2022towards} to obtain the desired result. Note that, having $\hat{x}_f$ as part of the hybrid state is not a problem. Indeed, it does not change at jumps from \eqref{eq:filteredStateEstimateJump} and, from \eqref{eq:filteredStateEstimateFlow}, it is an input-to-state stable system in cascade with the hybrid system used in \cite[Theorem 1]{petri2022towards}, see \cite[Section 4]{sontag2008input}.
 \hfill $\blacksquare$
%\color{black}

%The proof of Theorem~\ref{thm:LyapunovSolutionProposition} is omitted for space reasons and it follows from \cite[Proof of Theorem 1]{petri2022towards}, the fact that $\hat{x}_f$ does not change at jumps and because \eqref{eq:filteredStateEstimateFlow} is an input-to-state stable system in cascade with the hybrid system used in \cite[Theorem 1]{petri2022towards}, see \cite[Section 4]{sontag2008input}. 
%
%\textcolor{red}{I didn't write carefully the proof yet to be sure, but as Romain wrote on Slack, this should come automatically from ISS + cascade. I will try to check it carefully (its two measure ISS and hybrid system) in the next days, but I think there should not be problems. Do you suggest a paper to cite for this if we do not provide the proof? or Sontag one is ok?}

Theorem~\ref{thm:LyapunovSolutionProposition} ensures that the estimation errors and the monitoring variables of the nominal observer $e_1$ and $\eta_1$ converge to a neighborhood of the origin, whose ``size" depends on the $\mathcal{L}_\infty$ norm of $v$ and $w$, which is not surprising in view of Theorem~\ref{thm:observerTheorem}. However, Theorem~\ref{thm:LyapunovSolutionProposition} also guarantees that the state estimation error and the monitoring variable of the hybrid multi-observer $e_\sigma$ and $\eta_\sigma$ and also the filtered version of the estimation error, namely $e_f$, converge to the same neighborhood of the origin. Hence, the convergence of the (filtered) state estimate produced by the hybrid scheme is guaranteed despite the fact that the gains $L_k$ in \eqref{eq:ObserverAdditional} were freely selected.
%\textcolor{red}{Modify this comment: see Romain comments on v2, but how? }
%
%\color{blue}
Moreover, when $\varepsilon =1$, we have that $\eta_{\sigma(t,j)}(t,j) \leq \eta_1(t,j)$, for all $ (t,j)$ in the domain of the solution $q$. Therefore, the estimation performance of the hybrid multi-observer are always not worse than the nominal one, according to the considered performance cost. We will see in the next section that significant  performance improvements can be obtained in simulations.
%\color{black}

\section{NUMERICAL STUDY}\label{Simulations}
In this section, we compare the estimates generated by a nominal observer \eqref{eq:systeme_obs} and the associated hybrid multi-observer \eqref{eq:HybridSystemGeneral} with standard parameter values.
%
% we show the simulation results obtained when the hybrid multi-observer is used to estimate the states of the lithium-ion battery in \eqref{eq:systeme_state_space}. 

\subsection{System model}
\begin{figure}
	\centering
	\includegraphics[width=0.8\linewidth]{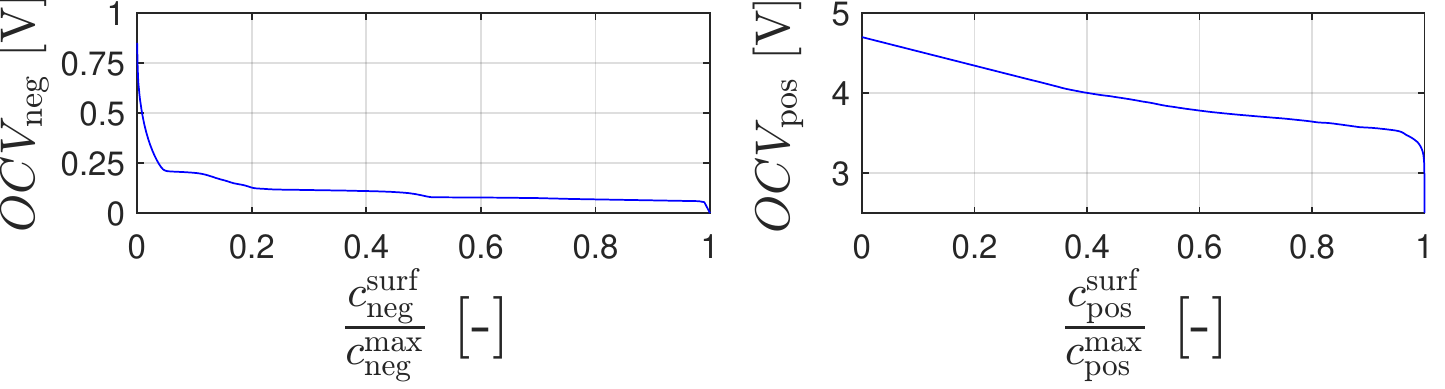}\\ 
	\vspace*{-0.5cm} \caption{OCV curves}
	\vspace*{-0.3cm}
	\label{Fig:OCV}
\end{figure}

We assume that each electrode is composed of $6$ samples with identical volumes. Consequently, $N_{\text{neg}} = N_{\text{pos}}=6$ and $n_x = N_{\text{neg}}-1 + N_{\text{pos}} = 11$. %We consider the set of parameters in \cite[Table~I]{blondel2018nonlinear}. 
We consider the parameters in Table~\ref{Tab:parameters} in the appendix.
%The disturbance input $v$ is taken equal to~$0$ and the matrix $E = [1, \dots, 1] \in \R^{7 \times 1}$. 
We take a measurement noise equal to~$0.05\sin(30t) \, V$, which has a reasonable frequency and signal versus noise ratio for embedded battery voltage measurements. The input $w$ in  \eqref{eq:systeme_state_space} is given by $w=0.05\sin(30t) + w_{2}(t)$ where $w_2$ is an additional term due to the input mismatch between the battery and its observer as clarified in the sequel. %The disturbance input $v$ in \eqref{eq:systeme_state_space} is used to model a mismatch in the current input to the battery, and its biased version known by the observer, see Section~\ref{InputCurrentSimulation}. For this purpose, we consider $E=B$. %Band, for the same reason, the matrix $E$ is taken equal to the matrix $B$. We will explain this in the next section. %To consider this frequency and this ratio signal versus noise is reasonable for embedded battery voltage measurements.  %\textcolor{red}{which is ???}. 
The considered OCV curves for the positive and the negative electrodes are shown in Fig.~\ref{Fig:OCV}, which satisfy Assumption~\ref{ASS:OCVassumption} and \cite[Assumption 2]{petri2022towards}.

%\textcolor{red}{ADD SOME COMMENTS ON Ev MODELLED AS THE MISMATCH IN THE CURRENT DUE TO THE BIAS}

\subsection{Input current}\label{InputCurrentSimulation}
\begin{figure}
	\centering
	\includegraphics[trim= 0cm 0cm 0cm 0cm, clip, width=0.78\linewidth]{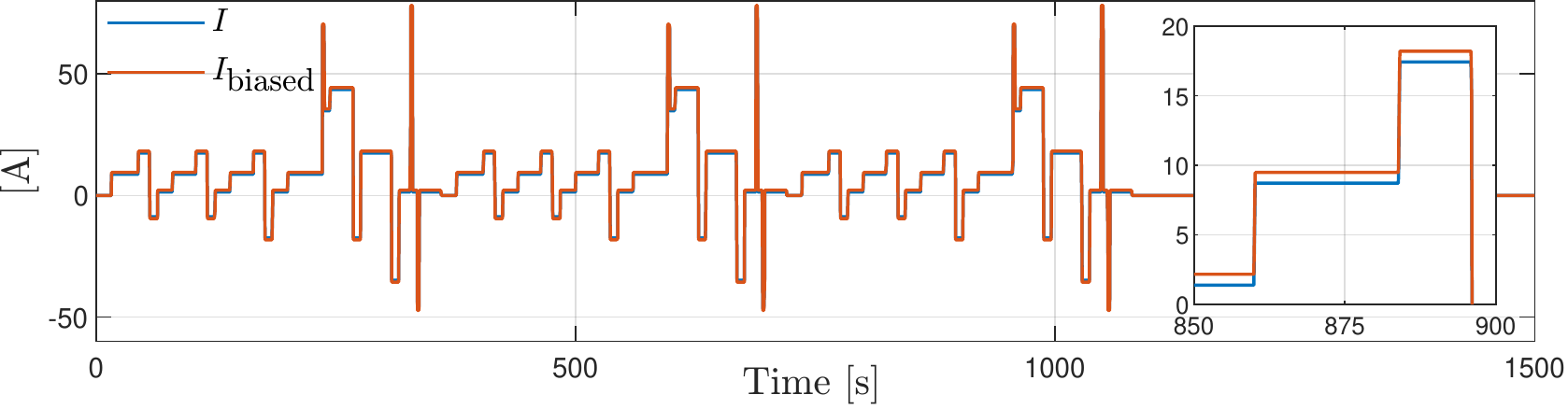}\\ 
	\vspace*{-0.3cm} \caption{Input current profile and its biased version available to the observer}
	\label{Fig:InputCurrent}
\end{figure}

The input $u$ is given by a Plug-in Hybrid Electrical Vehicles (PHEV) current profile \cite{belt2010battery}. %, which is typically used to test both the model and the observer. 
In practical applications, the observer usually only knows a biased version of the battery current. This bias is due to the precision of the sensor and its conditioning.
We therefore introduce $I_{\text{biased}}$ to denote the input $u$ known by the observer, which is given by
%
%To replicate this phenomenon, the observer input current we use in the simulations is given by
%$I_{\text{biased}}(t) =  \begin{cases}
%	0 \quad  &I(t) = 0\\
%	I(t) + 0.01\times \max \limits_{t^\star \in [0,t]}I(t^\star) \quad & I(t) \neq 0\\
%\end{cases}
%$, for all $t \geq 0$.
$I_{\text{biased}}(t) =  0$ when $I(t) = 0$,  $I_{\text{biased}}(t) = I(t) + 0.01 \max \limits_{t^\star \in [0,t]}|I(t^\star)|$ when $I(t) > 0$ and  $I_{\text{biased}}(t) = I(t) - 0.01 \max \limits_{t^\star \in [0,t]}|I(t^\star)|$ when $I(t) < 0$, for all $t \geq 0$. We consider a precision of $1 \%$ on the full scale for the current bias, which corresponds to a standard sensor. 
The PHEV current input $I$ and its biased version $I_{\text{biased}}$ are shown in Fig.~\ref{Fig:InputCurrent}. 
%
%\color{blue}
This mismatch in the current input of system and observer can be modeled using the disturbance input $v$ and the exogenous perturbation $w$ in \eqref{eq:systeme_state_space}. Indeed, the plant input $u = I = I_\text{biased} +v$, where $v$ is defined as $v := I - I_{\text{biased}}$. With the matrix $E$ equal to the matrix $B$, we obtain $\dot x = Ax + BI_{\text{biased}} + K + Bv$. Moreover, to model the input mismatch in the output map, we define $w_2 = g(I_{\text{biased}})-g(I)$, so that $w$ in \eqref{eq:systeme_state_space} is  $w=0.05\sin(30t)+g(I_{\text{biased}}) -g(I)$. %the input mismatch in the output map, i.e., $g(I) - g(I_{\text{biased}})$, can be included in the exogenous perturbation $w$.
%} % and $\dot{\hat x}_1 = A\hat x_1 + BI_{\text{biased}} + K + L_1(y-\hat y_1)$.  Note that this is only one possible option to model the input mismatch in the plant equation and other choices for $v$ and $E$ can be made.
%Moreover, we can use the additional perturbation $d$ in \eqref{eq:erreur} to model the mismatch in the output equations of system and observer due to the biased input. Indeed, defining $d := + L_1(g(I_{\text{biased}} + v) -g(I_{\text{biased}})) + \tilde{d} \in \R^{n_x}$, with $\tilde{d} \in \R^{n_x}$, we obtain \eqref{eq:erreur_avec_hypothese} with $\tilde{d}$ instead of $d$, and the results in this paper hold. 
%\color{black}
%\begin{figure}
%	\centering
%	\includegraphics[trim= 3.8cm 9.5cm 4.1cm 10.0cm, clip, width=0.5\linewidth]{./Figure/Current.pdf}\\ 
%	\caption{Input current profile and its biased version available to the observer}
%	\label{Fig:InputCurrent}
%\end{figure}
%

\subsection{Electrolyte dynamics}\label{ElectrolyteDynamicsSimulation}
To test the robustness of the estimation scheme, we consider a model of the electrolyte dynamics, as in \cite[Section IV.B]{plante2022multiple}, thereby relaxing item (\textit{ii}) in Assumption~\ref{ASS:singleSphereParticle}. Consequently, the battery output voltage becomes 
\begin{equation}
	y - \varrho_{\text{pos}} - \varrho_{\text{neg}} - \varrho_{\text{sep}},
\end{equation}
where $y$ is the battery output from \eqref{eq:systeme_state_space} and $\varrho_{\text{r}}$, with $r \in \{\text{pos, neg or sep}\}$, is the electrolyte diffusional overvoltage in the positive electrode, negative electrode or separator, which dynamics is given by $\dot \varrho_{\text{r}} = -\varrho_{\text{r}}/\varsigma_{1, \text{r}} + u\varsigma_{2, \text{r}} / \varsigma_{1, \text{r}}$, where $\varsigma_{1, \text{r}}$ and $\varsigma_{2, \text{r}}$ are the ionic diffusion time constant and ionic diffusion resistance in $r$. However, these electrolyte dynamics are ignored below when designing the nominal observer and the additional modes.

%The disturbance input $v$ is taken equal to~$0$ and the matrix $E = [1, \dots, 1] \in \R^{7 \times 1}$. We consider a measurement noise $w$ equal to~$0.05\sin(30t) \, V$ and $D =1$. 
%The considered OCV curves for the positive and the negative electrodes are shown in Fig.~\ref{Fig:OCV}, which satisfy Assumption~\ref{ASS:OCVassumption} and \cite[Assumption 2]{petri2022towards}.
%
%\begin{figure}
%	\centering
%		\includegraphics[trim= 3.3cm 12.05cm 3.7cm 12.4cm, clip, width=0.9\linewidth]{./Figure/OCV.pdf}\\ 
%	\caption{OCV curves \textcolor{red}{Reference??}}
%	\label{Fig:OCV}
%\end{figure}
%
%\textbf{
%\begin{figure}
%	\centering
%	\includegraphics[trim= 2.5cm 12.90cm 2.7cm 11.4cm, clip, width=0.95\linewidth]{./Figure/OCV3.pdf}\\ 
%	\caption{OCV curves for lithium-ion batteries \textcolor{red}{Reference??}}
%	\label{Fig:OCV}
%\end{figure}
%}
\subsection{Nominal observer}\label{NominalObserverResults}
\begin{figure}
	\centering
	\includegraphics[trim= 0cm 0.0cm 0cm 0cm, clip, width=0.76\linewidth]{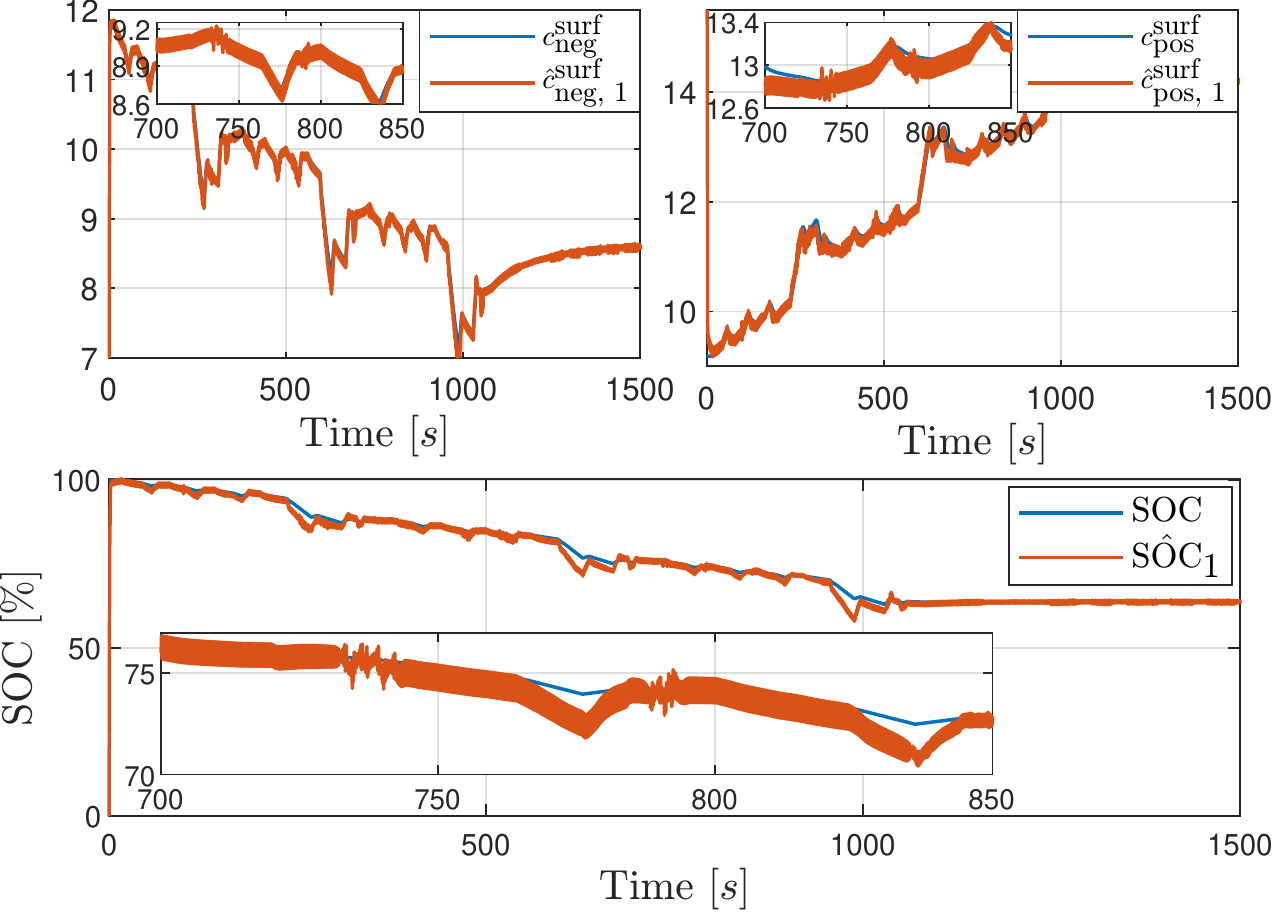}
	\vspace*{-0.3cm} \caption{Lithium concentrations at the surface of both electrodes $c_\text{neg}^\text{surf}$ (top figure left) and $c_\text{pos}^\text{surf}$ (top figure right), state of charge (SOC) (bottom figure). Reference system (blue), nominal observer (red).}
	\label{Fig:NominalResult}
\end{figure}

%\color{blue}
We now design the nominal observer in \eqref{eq:systeme_obs}. % using %and 
To test its efficiency, we design it with 
a smaller number of samples compared to the system model in~\eqref{eq:systeme_state_space}. In this way, a higher fidelity model is used to generate the output voltage. We thus select  $N_{\text{neg, obs}} = N_{\text{pos, obs}}=4$ and $n_{x, \text{obs}} = N_{\text{neg, obs}}-1 + N_{\text{pos, obs}} = 7$, while the battery model is $11+3$, where the $3$ additional dimensions are due to the electrolyte dynamics in Section~\ref{ElectrolyteDynamicsSimulation}. %In this way, the battery model used to generate the output voltage $y$ is more accurate. 
We then solve \eqref{eq:LMI_ThISSNominal} and we obtain $L_1 = (28.03, 27.78, 28.77, -45.54, -45.72, -44.78, -46.28)$.
%We then design the nominal observer in \eqref{eq:systeme_obs} and, solving \eqref{eq:LMI_ThISSNominal}, we obtain $L_1 = (28.03, 27.78, 28.77, -45.54, -45.72, -44.78, -46.28)$.
The system is initialized with a state of charge of $100 \%$, which corresponds to $x(0,0) = (11.75,11.75,11.75,11.75,11.75, 9.182, 9.182, 9.182, 9.182, $\\ $9.182, 9.182)$, %\footnote{All the signals involved in the hybrid multi-observer are defined on hybrid-time domains \cite[Definition 2.3]{goebel2012hybrid}, where the first argument is the continuous time $t$, while the second is the discrete time $j$.},
 while the nominal observer is initialized with a state of charge of $0 \%$, which corresponds to $\hat{x}_1(0,0) = (3.069, 3.069, 3.069, 23.01,23.01,23.01, $ $ 23.01)$. Therefore, the state of charge estimation error is initialized at $100 \ \%$, which is the largest possible initial estimation error. The electrolytes diffusional overvoltages are initialized at $\varrho_{\text{r}}(0,0) = 0$ for any $r \in \{\text{pos, neg, sep}\}$. 
The lithium surface concentrations, and their estimations, of both the negative and positive electrodes are shown in Fig.~\ref{Fig:NominalResult}, together with the state of charge and its estimate.  

The nominal observer has good performance in terms of speed of convergence, see Fig.~\ref{Fig:NominalResult}. Indeed, despite the large initial error for the SOC, %even when the state of charge estimation error is initialized at $100 \%$, which is the largest possible initial error, 
the nominal observer estimate converges fast to the actual SOC. However, the observer estimates is very sensitive to measurement noise, model mismatch and input bias, which impact the estimation performance especially when the estimation error reaches a neighborhood of the origin. Consequently, the hybrid multi-observer is designed in the next section with the aim of improving the estimation performance in terms of robustness to measurement noise, model mismatch and input bias, while preserving the fast convergence of the nominal observer. 
%\color{black}

\subsection{Hybrid multi-observer}
We design the multi-observer adding $N=3$ additional modes~\eqref{eq:ObserverAdditional} in parallel to the nominal observer. %These modes have the same structure as the nominal one, but different gains $L_k \in \R^{7\times 1}$, with $k \in \{2, 3,4\}$.
% The selection of the additional gains $L_k \in \R^{7\times 1}$, with $k \in \{2, 3,4\}$, is done based on the behavior of the nominal observer we obtained from simulation %In particular, during the transient, the speed of convergence of the nominal observer is good, even when we have the largest possible initial estimation error. However, the nominal observer is very sensitive to noise, which impacts the estimation performance especially when the estimation error reaches a neighborhood of the origin. 
% From Fig.~\ref{Fig:NominalResult} we deduce that the hybrid multi-observer has to be designed with the aim of improving the estimation performance in terms of robustness to noise and, thus, 
 %\color{blue}
 %and,
  Since small gains  typically help with respect to noise, we chose the additional gains smaller than the nominal one, even though they may not result in converging estimation errors. 
% \color{black}
%we chose the additional gains smaller than the nominal one.
In particular, we select $L_2 = L_1/10$, $L_3 = L_1/100$ and $L_4 = 0_{7 \times 1}$. %For these choice of gains, 
The gain $L_4 = 0_{7 \times 1}$ does not lead to a ``converging mode'' but it is the best choice to annihilate the measurement noise. %does not amplify the measurement noise, conversely to all other possible gain selections. Thus, it can be a good choice to improve the robustness of the observer. 
%}
	Simulations suggest that the SOC estimation error of the modes with $L_2$ and $L_3$ converge, while the one with $L_4$ does not. Note that, in the choice of the additional gains we exploited the complete freedom given in Section~\ref{HybridMultiObserver}. %Indeed, the gains are not designed using techniques that guarantee a convergence property of the estimation errors.

\subsection{Initialization and design parameters}
%The system is initialized with a state of charge of $100 \%$, which corresponds to $x(0,0) = (11.75,11.75,11.75,11.75,11.75, 9.182, 9.182, 9.182, 9.182, $\\ $9.182, 9.182)$\footnote{All the signals involved in the hybrid multi-observer are defined on hybrid-time domains \cite[Definition 2.3]{goebel2012hybrid}, where the first argument is the continuous time $t$, while the second is the discrete time $j$.}, while all the modes of the multi-observer are initialized with a state of charge of $0 \%$, which corresponds to $\hat{x}_k(0,0) = (3.069, 3.069, 3.069, 23.01,23.01,23.01, $ $ 23.01)$, for all $k \in \{1, \dots, 4\}$. Therefore, the state of charge estimation error is initialized at $100 \ \%$, which is the largest possible initial estimation error. The electrolytes diffusional overvoltages are initialized at $\varrho_{\text{r}}(0,0) = 0$ for any $r \in \{\text{pos, neg, sep}\}$. 
%The system is initialized with a state of charge of $100 \%$, as before in the simulations with only the nominal observer, while all the modes of the multi-observer are initialized with a state of charge of $0 \%$, which is the same initial condition we consider for the nominal observer in the simulation in Section~\ref{NominalObserverResults} and corresponds to $\hat{x}_k(0,0) = (3.069, 3.069, 3.069, 23.01,23.01,23.01, $ $ 23.01)$, for all $k \in \{1, \dots, 4\}$. The electrolytes diffusional overvoltages are initialized at $\varrho_{\text{r}}(0,0) = 0$ for any $r \in \{\text{pos, neg, sep}\}$. 

%\color{blue}
The state estimate of the additional modes, $\hat x_k$, with $k \in \{2, \dots, 4\}$ are initialized at the same value as $\hat x_1$ in Section~\ref{NominalObserverResults}.
%\color{black}
%
We select $\eta_1(0,0) = 1$ and $\eta_k(0,0) = 10$ for all $k \in \{2, 3, 4\}$,  $\sigma(0,0) = 1$ and $\hat{x}_f(0,0) = \hat{x}_{\sigma(0,0)}(0,0) = \hat{x}_1(0,0)$. This choice of initializing the nominal monitoring variable $\eta_1$ smaller than the monitoring variables of all the additional modes is because the transitory performance of the nominal observer is good and this choice, together with the initialization of $\sigma$ at the nominal observer, allows to select the nominal observer for some amount of time at the beginning of the simulation. 
We simulate the proposed hybrid multi-observer with $\nu = 0.005$, $\lambda_1 = 1$, $\lambda_2 = 0.005$,  $\varepsilon= 0.95$ and $\zeta = 3$. Note that, the condition $\nu \in (0,\alpha]$ in \cite[Proposition~1]{petri2022towards} is satisfied. Indeed, for the considered lithium-ion battery, we have $\alpha = 0.01$ in Theorem~\ref{thm:observerTheorem}.
%The norm of the state estimation error of the nominal observer, $|e_1|$, as well as the one of the hybrid multi-observer $|e_\sigma|$ both without and with resets are shown in Figure \textcolor{red}{Add the figure}, together with the SOC and its estimations (with nominal and both cases of the multi-observer) and the signal $\sigma$. 

\begin{figure}
	\centering
	\includegraphics[trim= 0cm 0.0cm 0cm 0cm, clip, width=0.85\linewidth]{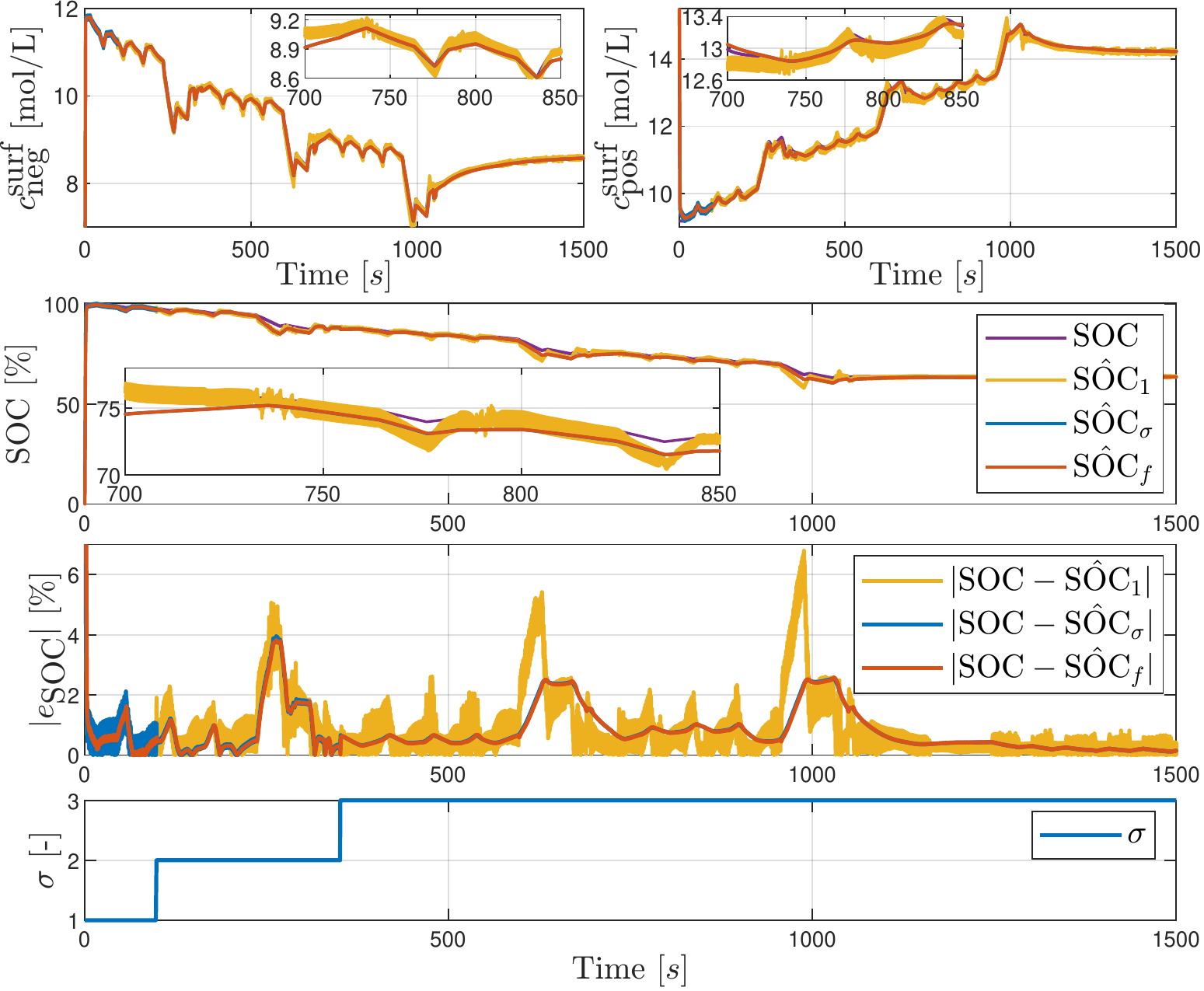}
	\vspace*{-0.2cm} 
	\caption{Lithium concentrations at the surface of both electrodes $c_\text{neg}^\text{surf}$ (top figure left) and $c_\text{pos}^\text{surf}$ (top figure right), state of charge (SOC) (second figure), norm of the state of charge estimation error (third figure) and $\sigma$ (bottom figure). Real system (purple), nominal (yellow), $\sigma$-estimate (blue), filtered estimate (red).}
	\label{Fig:Result}
\end{figure}

\subsection{Results} \label{SimulationResults} 
The lithium surface concentrations of both the negative and positive electrodes, namely $c_\text{neg}^\text{surf}$ and  $c_\text{pos}^\text{surf}$, together with their estimates using the nominal observer, the hybrid multi-observer and its filtered version are shown in Fig.~\ref{Fig:Result}. We recall that the lithium surface concentrations are elements of the system state and therefore Fig.~\ref{Fig:Result} shows that the hybrid multi-observer improve the state estimation performance compared to the nominal observer. Moreover, using \eqref{eq:SOC}, we obtain the state of charge (SOC) and its estimates with the nominal observer and the hybrid multi-observer (filtered and not) and, from these, we evaluate the norm of the state of charge estimation errors. The results are shown in Fig.~\ref{Fig:Result}, where we see that the state of charge estimate is improved, both on the averaged value and on the oscillations, using the hybrid multi-observer. 
The obtained performance improvement is commonly considered to be significant for this application.
%
%Fig.~\ref{Fig:Result} shows that the hybrid multi-observer improve the estimation performance compared to the nominal observer both in the case without and with resets. 
%
The last plot in Fig.~\ref{Fig:Result} represents the signal $\sigma$ which indicates the mode that is selected at every time instant.

To further evaluate the effectiveness of the proposed hybrid multi-observer, we have run $100$ simulations with different initial conditions. In particular, the initial state of charge estimate of all the modes of the multi-observer $\hat{\text{SOC}}_k(0,0)$, with $k \in \{1, \dots, 4\}$, were selected randomly in the interval $[0, 100] \%$, while the battery state of charge was always initialized at $\text{SOC}(0,0) = 100 \%$. We considered the same choice as before for all the design parameters and initial conditions of the monitoring variables $\eta_k$, with $k \in \{1, \dots, 4\}$, $\sigma$ and $\varrho_{\text{r}}$, with $r \in \{\text{pos, neg, sep}\}$.
To quantify the improvement brought by the hybrid multi-observer, we evaluate the mean absolute error (MAE) and the root mean square error (RMSE), averaged over all the simulations, on the SOC estimation error obtained with the nominal observer and the proposed hybrid multi-observer, filtered and not. The data collected are shown in Table~\ref{Tab:MAEandRMSE_random} for the whole simulation time $t \in [0, 1500]\,\text{s}$, during the transitory for $t \in [0, 150]\,\text{s}$ and after the transitory for $t \in [150, 1500]\,\text{s}$. 

\setlength{\tabcolsep}{5pt}
\setlength{\extrarowheight}{1pt}
\begin{table}[]
	%\scriptsize
	%\footnotesize
	\centering
%	\caption{Average over $100$ simulations with different $\hat{\textnormal{SOC}}_k(0,0)$, with $k \in \{1, \dots,4\}$,  of the MAE and RMSE of the SOC estimation error ($e_{\text{SOC}}$) for $t \in [0, 1500]\,$\lowercase{s} (tot), $t \in [0, 150]\,$\lowercase{s} (tran) and $t \in [150, 1500]\,$\lowercase{s} (end).  }
	\caption{Average MAE and RMSE of the SOC estimation error ($e_{\text{SOC}}$) for $t \in [0, 1500]\,$\lowercase{s} (tot), $t \in [0, 150]\,$\lowercase{s} (tran) and $t \in [150, 1500]\,$\lowercase{s} (end).}
\label{Tab:MAEandRMSE_random}
	\begin{tabular}{l|ccc}
		%\toprule
		& $e_{\text{SOC},1}$ & 	$e_{\text{SOC},\sigma}$& 	$e_{\text{SOC},f}$\\
		\hline
		$\text{MAE}_{\text{tot}} [\%]$& 0.82&\textbf{0.76} &\textbf{0.76} \\ 
		$\text{MAE}_{\text{tran}} [\%]$& 0.83 & \textbf{0.81}&0.84 \\ 
		$\text{MAE}_{\text{end}} [\%]$& 0.82&0.76 &\textbf{0.75} \\ 
		\hline
		$\text{RMSE}_{\text{tot}} [\%]$&1.65 &\textbf{1.47} &1.64 \\ 
		$\text{RMSE}_{\text{tran}} [\%]$&3.07 &\textbf{3.06} &3.74 \\ 
		$\text{RMSE}_{\text{end}} [\%]$& 1.29& 1.04&\textbf{1.02} \\ 
		%\bottomrule
	\end{tabular}
\end{table}

Table \ref{Tab:MAEandRMSE_random} shows that the hybrid multi-observer unfiltered improves the estimation performance, especially at large times as desired. Indeed, both the MAE and the RMSE are always smaller compared to the ones of the nominal observer. Moreover, the filtered version, even if during transient has worse performance compared to the nominal observer, after the transient the improvement is clear and, the performance can be also better than the corresponding unfiltered version.

\section{CONCLUSIONS}\label{conclusions}
We have applied and extended the hybrid multi-observer proposed in \cite{petri2022towards} to improve the estimation performance of the observer based on a polytopic approach designed in \cite{blondel2017observer} to estimate the lithium concentration of the electrodes of an electrochemical battery, which is directly related to the state of charge. %We have also extended the hybrid multi-observer in \cite{petri2022towards} adding a filtered version of its estimate, which allows to solve the possible issue of discontinuity produced by the switching and we have provided a stability theorem for the new hybrid system. Finally, s
Simulations based on standard model parameter values have illustrated the potential of this approach to improve the state of charge estimation performance. %are shown and the estimation performance improvement has been quantified in terms of MAE and RMSE of the state of charge estimation error. 

In future work, we plan to include uncertainties in the design parameters and apply the proposed approach to experimental data.
%\textcolor{red}{Write some future directions... as the online implementation, as suggested by Rev 3, comment 12.}
%\textcolor{red}{
%In future work, we would like to include uncertainties in the design parameters and design the hybrid multi-observer considering some different nominal observers for lithium-ion battery state estimate. Moreover, it will also be interesting to apply the proposed approach to a real battery.
%}

%\section*{APPENDIX}
\appendix

\subsection{Model description}\label{appendix_definitions}
The matrices and function definitions in \eqref{eq:systeme_state_space} are defined as 
\begin{equation}
	A:= \begin{pmatrix}
		\multicolumn{2}{c}{A_{2,\text{neg}}}\\
		A^\star & \text{diag}(A_\text{neg}, A_\text{pos}),\\
	\end{pmatrix}
\text{ with }  A^\star:= \begin{pmatrix} 
	\mu_{3,2}^\text{neg}\\
	0_{(n_x-2) \times 1}
\end{pmatrix},
	\label{eq:A_matrix}
\end{equation}
%\begin{equation}
%	A:= \begin{pmatrix}
%		\multicolumn{2}{c}{A_{2,\text{neg}}}\\
%		\multirow{2}{*}{\mu_{3,2}^\text{neg}\\
%			0_{(n_x-2) \times 1}&\text{diag}(A_\text{neg}, A_\text{pos})},\\
%		A^\star & \text{diag}(A_\text{neg}, A_\text{pos}),\\
%	\end{pmatrix}
%	\label{eq:A_matrix}
%\end{equation}
where $A_{2,\text{neg}}\in \R^{1\times n_x}$ is 
%and $A^\star\in \R^{(n_x - 1 )\times 1}$ are
 defined as
$$
%A_{2,\text{neg}}^\top:= 
%\begin{pmatrix} 
%	\tilde{\nu}_{2}^{\text{neg}} \\ \tilde{\tilde{\nu}}_2^{\text{neg}} \\ \nu_{2,1,4}^{\text{neg,neg}} \\\vdots \\ \nu_{2,1,N_{\text{neg}}}^{\text{neg,neg}} \\
%	\nu_{2,1,1}^{\text{neg,pos}} \\ \vdots \\  \nu_{2,1,N_{\text{pos}}}^{\text{neg,pos}}
%\end{pmatrix}
\setlength\arraycolsep{1pt}
A_{2,\text{neg}}:= 
\begin{pmatrix} 
	\tilde{\nu}_{2}^{\text{neg}} & \tilde{\tilde{\nu}}_2^{\text{neg}} & \nu_{2,1,4}^{\text{neg,neg}} &\dots & \nu_{2,1,N_{\text{neg}}}^{\text{neg,neg}} &
	\nu_{2,1,1}^{\text{neg,pos}}& \dots &  \nu_{2,1,N_{\text{pos}}}^{\text{neg,pos}}
\end{pmatrix}, $$
%$$
% A^\star:= \begin{pmatrix} 
%	\mu_{3,2}^\text{neg}\\
%	0_{(n_x-2) \times 1}
%\end{pmatrix} ,
%$$  
while $A_{\text{neg}}\in \R^{N_{\text{neg}-2} \times N_{\text{neg}-2}}$, resp.
$A_{\text{pos}}\in \R^{N_{\text{pos}} \times N_{\text{pos}}}$,
as
%$$
%\begin{aligned}
%	A_{\text{neg}}&:= \underline{\text{diag}}(\mu_{i,i-1}^{\text{neg}}) + \text{diag}(\tilde{\mu_i}^\text{neg}) + \overline{\text{diag}}(\mu_{i,i}^\text{neg})    \\
%	A_{\text{pos}}&:= \underline{\text{diag}}(\mu_{i,i-1}^{\text{pos}}) + \text{diag}(\tilde{\mu_i}^\text{pos}) + \overline{\text{diag}}(\mu_{i,i}^\text{pos})
%\end{aligned}
%$$
%\begin{aligned}
	$A_{\text{neg}}:= \underline{\text{diag}}(\mu_{i,i-1}^{\text{neg}}) + \text{diag}(\tilde{\mu_i}^\text{neg}) + \overline{\text{diag}}(\mu_{i,i}^\text{neg})$, $   
	A_{\text{pos}}:= \underline{\text{diag}}(\mu_{i,i-1}^{\text{pos}}) + \text{diag}(\tilde{\mu_i}^\text{pos}) + \overline{\text{diag}}(\mu_{i,i}^\text{pos})$,
%\end{aligned}
for $ i \in \{3, \dots, N_{\text{neg}}\}$, 
resp., for $ i \in \{1, \dots, N_{\text{pos}}\}$, where $\underline{\text{diag}}$ denotes the lower diagonal, $\overline{\text{diag}}$ denotes the upper diagonal, 	$\tilde{\nu}_{2}^{\text{neg}}:= \nu_{2,1,2}^{\text{neg,neg}}-\mu_{2,1}^\text{neg}-\mu_{2,2}^\text{neg},$ $
\tilde{\tilde{\nu}}_2^{\text{neg}}:=  \nu_{2,1,3}^{\text{neg,neg}} + \mu_{2,2}^\text{neg}$ $ \mu_{i,j}^s:= \frac{D_{s}}{V_i^{s}}\frac{S_j^{s}}{r_{j+1}- r_j},$ 
$\tilde{\mu}_i^s:= -\mu_{i,i-1}^s-\mu_{i,i}^s$,
$\nu_{i,j,z}^{s,s'}:= \mu_{i,j}^s\beta_z^{s'},$ %$D_s$ is the lithium diffusion coefficient, $V_i^s$ is the volume of the sample $i$, $S_i^s$ is the external surface of the sample $i$ and $r_i$ is its radial coordinate, 
\( \beta_i^{\text{neg}} := -\frac{V_i^{\text{neg}}}{V_1^{\text{neg}}}\), 
$\beta_i^{\text{pos}} := -\frac{\alpha_{\text{pos}}V_i^{\text{pos}}}{\alpha_{\text{neg}} V_1^{\text{neg}}}$,  \( \alpha_{\text{s}}:= \frac{F}{3600} \frac{\varepsilon_{\text{s}}\mathcal{A}_{\text{cell}}d_{\text{s}}}{V_{\text{tot}}^{\text{s}}}\), %$\varepsilon_s$ is the volume fraction, $F$ is the Faraday's constant, $\cal{A}_\text{cell}$ is the cell area, $d_s$ is thickness of the electrode,
$D_s$, $V_i^s$, $S_i^s$, $r_i$, $\varepsilon_s$, $F$, $\cal{A}_\text{cell}$ and $d_s$ are defined in Table~\ref{Tab:parameters},
for any $ i, j,z \in \{1,...., N_{s}\}, s,s' \in \{\text{neg},\text{pos}\}$. 
The matrix $B$ is defined as 
%\begin{equation}
%	B :=(
%	0_{(N_{Neg-2})\times 1} \quad 
%	- \bar{K}_I^{\text{neg}} \quad
%	0_{(N_{pos-1})\times 1} \quad
%	\bar{K}_I^{\text{pos}} \quad
%	)^\top, %\in \R^{n_x\times 1}
%	\label{eq:B}
%\end{equation}
%\begin{equation}
	$B :=(
	0_{(N_{Neg-2})\times 1} \
	- \bar{K}_I^{\text{neg}} \
	0_{(N_{pos-1})\times 1} \
	\bar{K}_I^{\text{pos}} \
	)^\top,$ %\in \R^{n_x\times 1}
	%\label{eq:B}
%\end{equation}
with \( \bar{K}_I^{\text{s}} := - \frac{S_{\text{tot}}^{\text{s}}}{V_{1}^{\text{s}}a_{\text{s}}F \mathcal{A}_{\text{cell}}d_{\text{s}}}  \),
where $a_s:= 3\varepsilon_s/ R_s$, $R_s$ is the radius and $s \in \{\text{neg}, \text{pos}\}$.
The matrix $K$ is defined as 
%\begin{equation} 
%$	K :=(-	\mu_{2,1}^{\text{neg}}\bar{K} \quad  0_{(N-1)\times 1} )^\top,$ %\in \R^{n_x\times 1},
%\label{eq:K}
%\end{equation}
%\begin{equation} 
$	K :=(-	\mu_{2,1}^{\text{neg}}\bar{K} \  0_{(N-1)\times 1} )^\top,$ %\in \R^{n_x\times 1},
	%\label{eq:K}
%\end{equation}
with \( \bar{K} :=  \frac{Q_{Li}}{V_1^{\text{neg}}\alpha_{\text{neg}}}  \), where $Q_{Li}$ is the quantity of lithium in the solid phase.
Finally, 
%\begin{equation}
%	g(u):= g_1(u) + g_2(u)+g_3(u),
%\end{equation} 
%\begin{equation}
	$g(u):= g_1(u) + g_2(u)+g_3(u),$
%\end{equation} 
with 
%$$
%\begin{aligned}
%	g_1(u) &:= 2\frac{RT}{F}\text{Argsh}\Big(\frac{-R_{\text{pos}}}{6 \epsilon_{\text{pos}} j_0^{\text{pos}} \mathcal{A}_{cell} d_{\text{pos}}}u \Big),
%	\\
%	g_2(u)& := -2\frac{RT}{F}\text{Argsh}\Big(\frac{R_{\text{neg}}}{6 \epsilon_{\text{neg}} j_0^{\text{neg}} \mathcal{A}_{cell} d_{\text{neg}}}u \Big),
%	\\
%	g_3(u) &:= - \Big(\frac{1}{2\mathcal{A}_{cell}}\Big(\frac{d_{\text{neg}}}{\sigma_{\text{neg}}}+ \frac{d_{\text{pos}}}{\sigma_{\text{pos}}} \Big) + \Omega_{\text{add}} \Big )u,
%\end{aligned}	
%$$
%
%\begin{aligned}
	$g_1(u) := 2\frac{RT}{F}\text{Argsh}\Big(\frac{-R_{\text{pos}}}{6 \epsilon_{\text{pos}} j_0^{\text{pos}} \mathcal{A}_{cell} d_{\text{pos}}}u \Big),$
$	g_2(u) := -2\frac{RT}{F}\text{Argsh}\Big(\frac{R_{\text{neg}}}{6 \epsilon_{\text{neg}} j_0^{\text{neg}} \mathcal{A}_{cell} d_{\text{neg}}}u \Big),$	
$	g_3(u) := - \Big(\frac{1}{2\mathcal{A}_{cell}}\Big(\frac{d_{\text{neg}}}{\sigma_{\text{neg}}}+ \frac{d_{\text{pos}}}{\sigma_{\text{pos}}} \Big) + \Omega_{\text{add}} \Big )u,$
%\end{aligned}	
where $\text{Argsh}(\xi) = \text{ln}(\xi + \sqrt{\xi^2 +1})$ for any $\xi \in \mathbb{R} $. 

%\subsection{Values}\label{appendix_parameters}
%The battery parameters used in the simulations are given in Table~\ref{Tab:parameters}.

\setlength{\tabcolsep}{4pt}
\begin{table}[]
		\scriptsize
	%\footnotesize
	\centering
	\caption{Physical parameters of the electrochemical model}
	\label{Tab:parameters}
	\begin{tabular}{lll}% {|l|l|l|l|l|l|l|l|}
		\toprule
		$\mathcal{A}_{\text{cell}}$ & Cell area [$m^2$] & $1.0452$ \\
		$F$ & Faraday's constant [$C/mol$] &  $96485$  \\
		$R$ & Gas constant [$J/K/mol$] & $8.3145$ \\
		$T$ & Temperature [$K$] & $298.15$\\
		$N$                                         & Order of the model [-]           & $7$ \\
		$d_{\text{pos}}$                                       & Thickness
		of the positive electrode [$\mu m$]         & $36$ \\
		$d_{\text{neg}}$                                       & Thickness
		of the negative electrode  [$\mu m$] & $50$\\
		$D_{\text{pos}}$                     & Lithium diffusion coefficient  [$m^2/s$]     & \multicolumn{1}{l}{$3.723\times10^{-16}$}  \\
		$D_{\text{neg}}$                     & Lithium diffusion coefficient e  [$m^2/s$]     & \multicolumn{1}{l}{$2\times10^{-16}$}\\
		
		%	$ c_{{0},pos} $ &Lithium concentration at SOC = 0\% $[mol.L^{-1}]$ & $23.01$\\
		$ c_{{0},pos} $ &Lithium concentration at SOC = 0\% & $23.01$\\
		& $[mol.L^{-1}]$ &\\
		%	$ c_{{0},neg} $& Lithium concentration at SOC = 0\% $[mol.L^{-1}]$ & $3.069$\\
		$ c_{{0},neg} $& Lithium concentration at SOC = 0\% & $3.167$\\
		& $[mol.L^{-1}]$& \\
		%	$ c_{{100},pos} $&Lithium concentration at SOC = 100\% $[mol.L^{-1}]$ & $9.182$\\
		$ c_{{100},pos} $&Lithium concentration at SOC = 100\% & $9.182$\\
		&$[mol.L^{-1}]$&\\
		%$ c_{{100},neg} $ &Lithium concentration at SOC = 100\% $[mol.L^{-1}]$ & $11.75$\\
		$ c_{{100},neg} $ &Lithium concentration at SOC = 100\% & $11.75$\\
		&$[mol.L^{-1}]$&\\
		
		$c_{\text{max},\text{pos}}$ & Maximum concentration [$mol.L^{-1}$]   &
		\multicolumn{1}{l}{$23.9$}\\
		$c_{\text{max},\text{neg}}$ & Maximum concentration [$mol.L^{-1}$]   &
		\multicolumn{1}{l}{$16.1$}\\
		
		$\sigma_{\text{pos}}$ & Electronic conductivity [$S/m$]       &
		\multicolumn{1}{l}{$10$}\\
		$\sigma_{\text{neg}}$ & Electronic conductivity [$S/m$]       &
		\multicolumn{1}{l}{$100$}\\
		%	\textcolor{red}{	$\sigma_{\text{pos},\text{e}}$} & Ionic conductivity [$S/m$]       &
		%	\multicolumn{1}{l}{$0.6329$}\\
		%\textcolor{red}{	$\sigma_{\text{neg},\text{e}}$} & Ionic conductivity [$S/m$]       &
		%	\multicolumn{1}{l}{$0.6329$}\\
		
		$R_{\text{pos}}$                                        & Particle
		radius  [$\mu m$]         & \multicolumn{1}{l}{$1$}\\
		$R_{\text{neg}}$                                        & Particle
		radius  [$\mu m$]         & \multicolumn{1}{l}{$1$}\\
		$j_{0,{\text{pos}}}$                 & Exchange current density
		[$A/m^2$]     & $0.5417$ \\
		$j_{0,{\text{neg}}}$                 & Exchange current density 
		[$A/m^2$]     &  $0.75$ \\
		$\varepsilon_{\text{pos}}$&  Volume fraction of the material & \\
		& within the positive  electrode  [-] & $0.5$ \\
		$\varepsilon_{\text{neg}}$ &  Volume fraction of the material \\
		& within the negative electrode  [-] & $0.58$ \\
		$Q_{\text{Li}}$ & Lithium quantity in cell solid phases $[Ah]$ & $14.8318$ \\
		$Q_{cell}$ & Cell capacity [$Ah$] & $6.9725$ \\
		$\Omega_{\text{add}}$     & Additional resistivity $[\Omega]$  & $0$\\
		$\varsigma_{1, \text{pos}}$ & Ionic diffusion time constant $[s]$ & $13.0$\\
		$\varsigma_{1, \text{neg}}$ & Ionic diffusion time constant $[s]$ & $17.3$\\
		$\varsigma_{1, \text{sep}}$ & Ionic diffusion time constant of separator$[s]$ & $12.3$\\
		$\varsigma_{2, \text{pos}}$ & Ionic diffusion resistance $[\mu \Omega]$ & $153.9$\\
		$\varsigma_{2, \text{neg}}$ & Ionic diffusion resistance $[\mu \Omega]$ & $209.5$\\
		$\varsigma_{2, \text{sep}}$ & Ionic diffusion resistance of separator $[\mu \Omega]$ & $115.1$\\
		\bottomrule
	\end{tabular}
\end{table}

\bibliography{bibliography}

\end{document}